\documentclass[12pt, letterpaper]{JHEP3}
\usepackage{amsmath}
\usepackage{amsmath}
\usepackage{amssymb}
\usepackage{cite}
\usepackage{graphicx}
\usepackage{subfigure}

\newcommand{\comment}[1]{}

\title{Saturating the holographic entropy bound}

\author{Raphael Bousso$^{a,b,c}$, Ben Freivogel$^{a,b}$ and Stefan Leichenauer$^{a,b}$\\ \\
  $^a$ Center for Theoretical Physics, Department of Physics\\
\ \  University of California, Berkeley, CA 94720-7300, U.S.A.\\
$^b$  Lawrence Berkeley National Laboratory, Berkeley, CA 94720-8162,
  U.S.A.\\
$^c$  Institute for the Physics and Mathematics of the Universe,\\ 
\ \ University of Tokyo,
5-1-5 Kashiwa-no-Ha, Kashiwa City, Chiba 277-8568, Japan}

\abstract{The covariant entropy bound states that the entropy, $S$, of matter on a light-sheet cannot exceed a quarter of its initial area, $A$, in Planck units.  The gravitational entropy of black holes saturates this inequality.  The entropy of matter systems, however, falls short of saturating the bound in known examples.  This puzzling gap has led to speculation that a much stronger bound, $S\lesssim A^{3/4}$, may hold true.  In this note, we exhibit light-sheets whose entropy exceeds $A^{3/4}$ by arbitrarily large factors.  In open FRW universes, such light-sheets contain the entropy visible in the sky; in the limit of early curvature domination, the covariant bound can be saturated but not violated.  As a corollary, we find that the maximum observable matter and radiation entropy in universes with positive (negative) cosmological constant is of order $\Lambda^{-1}$ ($\Lambda^{-2}$), and not $|\Lambda|^{-3/4}$ as had hitherto been believed.  Our results strengthen the evidence for the covariant entropy bound, while showing that the stronger bound $S\lesssim A^{3/4}$ is not universally valid.  We conjecture that the stronger bound does hold for static, weakly gravitating systems.}

\begin{document}

\section{Introduction}

\paragraph{Covariant entropy bound}
The covariant entropy bound~\cite{CEB1} (see Ref.~\cite{RMP} for a review) establishes a general relation between quantum information and classical geometry: The entropy of matter on a light-sheet $L$ (a non-expanding null hypersurface) orthogonal to a spatial surface $B$ cannot exceed the surface area $A$, measured in Planck units:
\begin{equation}
S[L(B)]\leq \frac{A(B)}{4}~.
\label{eq-ceb}
\end{equation}
This holds for arbitrary spacelike surfaces $B$ of codimension two, open or closed, in any spacetime satisfying Einstein's equation with physically reasonable matter.  It implies that the entropy in the past of any event cannot exceed half of the maximum area of the past light-cone (which is finite in cosmological spacetimes, for example).

The covariant bound appears to be rather rigid.  Several other entropy bounds of the form $S\leq A/4$ have been formulated, which do not involve light-sheets or do not impose the condition $\theta\leq 0$ on the expansion of the geodesics generating the light-sheet.  A simple example is the ``spacelike entropy bound'', the claim that the entropy in any volume of space is less than the area of its boundary.  One finds that each of these bounds can be violated by arbitrarily large factors~\cite{FisSus98,KalLin99,RMP} in perfectly physical spacetimes.  Meanwhile, no counter-examples to the covariant entropy bound have been found~\cite{RMP}, and sufficient conditions have been identified that guarantee the bound's validity in a large class of spacetimes~\cite{FMW,BouFla03}.

From the viewpoint of local field theory, the covariant entropy bound is surprising, since one expects entropy to scale with volume.  The holographic principle, in its most general form, is the conjecture that this surprising relation between geometry and information must have a fundamental origin, in a theory of quantum gravity~\cite{CEB2}.  This expectation has been borne out in a special case, by the AdS/CFT correspondence~\cite{Mal97}.  The number of binary CFT degrees of freedom necessary to describe an AdS region of (sufficiently large) surface area $A$ is indeed of order $A$~\cite{SusWit98}.

But the covariant entropy bound applies much more broadly.  It holds in cosmological and other highly dynamical spacetimes that lack a fundamental quantum gravitational description, such as the interior of black holes.  In this general setting, one expects that the bound will constrain how a quantum gravity theory should be formulated and how spacetime should arise from it.  Light-sheets on which the entropy is {\em equal\/} to the area seem poised to play a distinguished role in the emergence of a classical geometry.  Therefore, matter systems that saturate the covariant entropy bound are of special interest.  

Remarkably, no explicit examples of such systems have been found to date. Of course, the Bekenstein-Hawking entropy of an event horizon is equal to the horizon area, $S_{\rm BH}=A/4$, and so, in a sense,\footnote{The black hole horizon {\em is\/} a light-sheet, which is crossed by the matter that created or fell into the black hole.  In order to say that a black holes saturates the covariant bound, one would like to view the black hole as a kind of matter object whose worldline crosses a different light-sheet of initial area $A_{\rm hor}$.  This is possible if one regards the stretched horizon as a timelike boundary of the black hole~\cite{SusTho93} and one terminates the light-sheet there.} it saturates the bound.  However, it is striking that no material objects, with ordinary, non-Bekenstein-Hawking entropy, are known to saturate the bound.  

\paragraph{A stronger bound?}
In fact, it would appear that systems made from ordinary matter obey the far stronger bound
\begin{equation}
S\lesssim A^{3/4}~,
\label{eq-a34}
\end{equation}
falling short of the holographic bound by a factor of $A^{1/4}$ in Planck units, an enormous factor for macroscopic systems.  Consider, for example, a spherical box of radius $R$ filled with radiation at temperature $T$.  By increasing $T$, one can increase the entropy, $S\sim R^3T^3$.  However, $T$ is bounded from above by the requirement that the box should not collapse into a black hole: $E\sim R^3T^4\lesssim R$.  The largest radiation entropy is attained at the threshold of collapse, when $S\sim R^{3/2}$, saturating the bound (\ref{eq-a34}) but, in the semiclassical regime $A\gg 1$, falling far short of the holographic bound (\ref{eq-ceb}).  

Another example that supports the stronger bound (\ref{eq-a34}) obtains in cosmology.  Consider the past light-cone of an observer at the time $t_E$ in a radiation-dominated flat FRW universe with vanishing cosmological constant.  If one follows the light-cone towards the past, its cross-sectional area initially expands, but then contracts until it vanishes at the Big Bang.  The sphere of maximum area, $A_{\rm AH}\sim t_E^2$, can be regarded as the origin of {\em two\/} light-sheets, one past-directed and one future-directed, which together form the past light-cone.   The covariant bound states that the entropy on each light-sheet must be less than $A_{\rm AH}/4$.   The actual entropy on each light-sheet can be estimated by noting that the past light-cone has comoving size comparable to the volume enclosed by the Hubble horizon at the time $t_E$.  Since the evolution is adiabatic, the entropy on the past light-cone is the same as the entropy within the Hubble radius at the time $t_E$.  The proper energy density of radiation at this time is $\rho_{\rm rad}\sim t_E^{-2}$, and the proper entropy density is $s\sim \rho_{\rm rad}^{3/4}\sim t_E^{-3/2}$.  The proper horizon volume is $\sim t_E^3$, so the total entropy on the light-sheet is $S\sim t^{3/2}\sim A_{\rm AH}^{3/4}$.  Again one finds the bound (\ref{eq-a34}) approximately saturated, but the holographic bound far from saturated, with a factor $A^{1/4}$ to spare.

There is no contradiction here: the holographic bound is just an inequality, and it is not surprising that many systems fall far short of saturating it.  In most cases the ratio $S/A$ is even smaller than in the above examples: consider, for example, a sphere surrounding a region with vanishing entropy, such as empty space or a crystal at zero temperature.  What is intriguing, however, is that it appears to be hard to exceed the ratio $S/A\sim A^{-1/4}$ attained by the above two examples.  Can we find ordinary matter systems whose entropy comes close to saturating the covariant entropy bound?   Or is the holographic bound needlessly lenient?

If there truly was a universal bound of the form (\ref{eq-a34}) for matter, we would be forced to reconsider the significance of the covariant entropy bound and the holographic principle.  What importance could we ascribe to an upper bound that is far from saturated in all known examples?  Perhaps it is the quantity $A^{3/4}$ that truly governs the information content in quantum gravity?  

In this paper, we rule out this possibility, and we reaffirm the fundamental stature of the covariant entropy bound.  We will find simple examples in which the covariant entropy bound is indeed saturated, and an even larger class of examples where the entropy on a light-sheet exceeds the stronger bound $A^{3/4}$ by arbitrarily large factors.

\paragraph{Relation to previous work}
A number of works have considered systems that may violate the naive bound $S\lesssim A^{3/4}$ or even saturate the holographic bound $S\leq A/4$.  

Prior to the covariant bound, Fischler and Susskind~\cite{FisSus98} proposed a cosmological holographic bound $S\leq A/4$ on the future light-cone of a point on the Big Bang, where $A$ is the area of the light-cone at some time $t$, and $S$ is the entropy on the portion of the light-cone below $t$.  In flat or open universes, this light-cone is an allowed light-sheet off of any of its cross-sectional surfaces, so the following example~\cite{FisSus98} applies also to the covariant entropy bound.  In a universe filled with a maximally stiff fluid ($p=\rho$), one finds $S/A\sim \sigma(t)$, where $\sigma$ is the comoving entropy density.  The area of the light-cone grows with time, so $A^{1/4}\sigma(t)$ will eventually exceed unity, and the naive bound $S\lesssim A^{3/4}$ will be violated, assuming that such a fluid can carry entropy and do so adiabatically ($\sigma=$const).  Moreover, assuming that one can achieve $\sigma\sim 1$, the fluid will saturate the covariant bound for arbitrarily large area $A$.  Neither of these assumptions, however, has yet been justified in a microscopic model of such matter.  In our examples below, the entropy is that of ordinary pressureless particles or radiation, and thus can be explicitly computed rather than posited.

In Ref.~\cite{CEB1}, a shell collapsing onto a black hole was considered and shown to {\em obey\/} the covariant bound.  Under a number of assumptions that erred on the side of larger entropy, this shell was seen to saturate the bound on a light-sheet well inside a black hole.  No attempt was made to show that all the assumptions can in fact be satisfied, and it remains unclear whether they can.  The constructions given in the present paper, by contrast, are completely explicit.  Among them, the collapsing dust ball studied in Sec.~\ref{sec-ball} is perhaps the example most similar to (yet definitely distinct from) the collapsing shell of Ref.~\cite{CEB1}.

``Monsters''~\cite{HsuRee07,HsuRee09} are matter configurations that violate the {\em spacelike\/} entropy bound, adding to a multitude of counterexamples~\cite{RMP}.  But monsters do obey the covariant bound, since the light-sheets off of their boundary surface are truncated by black hole singularities.  No evidence has been presented that the stronger inequality $S\lesssim A^{3/4}$ is violated on these light-sheets.

The scaling of entropy like $\Lambda^{-1}$ ($\Lambda^{-2}$) in open FRW universes with positive (negative) cosmological constant was noted earlier~\cite{BouLei09} in the context of the causal entropic principle~\cite{Bou06,BouHar07}.

\paragraph{Outline} In Sec.~\ref{sec-main}, we will study open FRW cosmologies with zero, positive, or negative cosmological constant.  We will identify light-sheets with entropy $S\gg A^{3/4}$, and we will show that in the limit of early curvature domination, the holographic bound can be saturated: $S\to A$.  

In Sec.~\ref{sec-patch} we will consider the past light-cone of an observer in such universes.  We will show that the observer's sky can be filled with so much entropy as to saturate the covariant bound.  This shows that the results of Sec.~\ref{sec-main} correspond naturally to directly observable situations.  As a corollary, we will find that the observable entropy in the presence of a cosmological constant can exceed the naive bound $|\Lambda|^{-3/4}$, and can become as large as $\Lambda^{-1}$ for $\Lambda>0$, and $\Lambda^{-2}$ for $\Lambda<0$.  

In Sec.~\ref{sec-ball}, we will show that light-sheets with $S\gg A^{3/4}$ exist not only in open universes.  We will demonstrate that such light-sheets can be actively produced by setting up a collapsing ball of pressureless matter with a particular velocity distribution.  In Sec.~\ref{sec-cd}, we go further and show that a light-sheet with $S\gg A^{3/4}$ can be both set up and observed by a single observer.  One relevant example arises when a black hole is slowly fed with quanta not much smaller than its own radius.

In Sec.~\ref{sec-conjecture} we will assess the possible role of the bound $S\lesssim A^{3/4}$.  We will briefly explore the conjecture that this bound holds for static, weakly gravitating systems.  The appendix will discuss the application of our results to the generalized covariant entropy bound~\cite{FMW}.

\section{Saturating the holographic entropy bound in cosmology}
\label{sec-main}

In this section, we will identify an interesting class of light-sheets arising in open FRW universes.  We will show that the entropy on these light-sheets can exceed the naive bound, $S\lesssim A^{3/4}$, by arbitrarily large factors.  We will also show that the holographic bound, $S\sim A$, can be saturated (but not violated), with ordinary matter or radiation, for arbitrarily large $A$.

Before we begin our detailed description of these light-sheets, we should, for completeness, mention a class of counterexamples to $S\lesssim A^{3/4}$ which we will not discuss in much detail.  In a flat radiation-dominated FRW universe, consider spheres at very large radius at some fixed time, $t$.  Such spheres possess two past-directed light-sheets, ingoing and outgoing.  In the large radius limit, both satisfy
\begin{equation}
\frac{S}{A}=t^{-1/2}~.
\label{eq-trivial}
\end{equation}
The key point is that the right-hand side is independently of $A$.  For sufficiently large radius, one has $A^{1/4}\gg t^{3/2}$ and the naive bound $S\lesssim A^{3/4}$ is violated. The geometric reason for this behavior is that the light-sheets of large spheres are ``truncated light-cones''~\cite{FisSus98} that end on the Big Bang while their comoving radius is still large.  Thus their comoving volume is a shell whose width is independent of $A$ and whose comoving volume and radiation entropy grows in proportion to $A$.  Similar light-sheets exist for pressureless matter, and they exist also in open and closed FRW universes.  To our knowledge, Eq.~(\ref{eq-trivial}) has never been held up as a counterexample to Eq.~(\ref{eq-a34}), though it is a perfectly valid one.

The counterexamples we will present in this section are closely related but arise only in open FRW universes.  They have the advantage that we will not need to require that the initial sphere be far outside the particle horizon.  This property, in turn, will be key to our demonstration, in Sec.~\ref{sec-patch}, that the maximum entropy in spacetimes with positive (negative) cosmological constant is $\Lambda^{-1}$ ($\Lambda^{-2}$).

\subsection{Light-sheets in open FRW universes}

We consider spatially open FRW universes, with metric
\begin{equation}
ds^2 = -dt^2+a(t)^2 dH_3^2~,
\label{eq-frwmetric}
\end{equation}
where
\begin{equation}
dH_3^2=d\xi^2 + \sinh^2\xi \, d\Omega^2
\end{equation}
is the metric on the unit three-hyperboloid.  For later use we note that the area of a sphere of radius $\xi$ on the unit three-hyperboloid is $A_c=4\pi \sinh^2 \xi$; the enclosed volume is $V_c(\xi)= \pi[\sinh(2\xi)-2\xi]$.   The physical area and volume are related to these comoving quantities by factors of $a^2$ and $a^3$, respectively.

\begin{figure}[ht]
\centering
\subfigure[]{
\includegraphics[width=1.5 in]{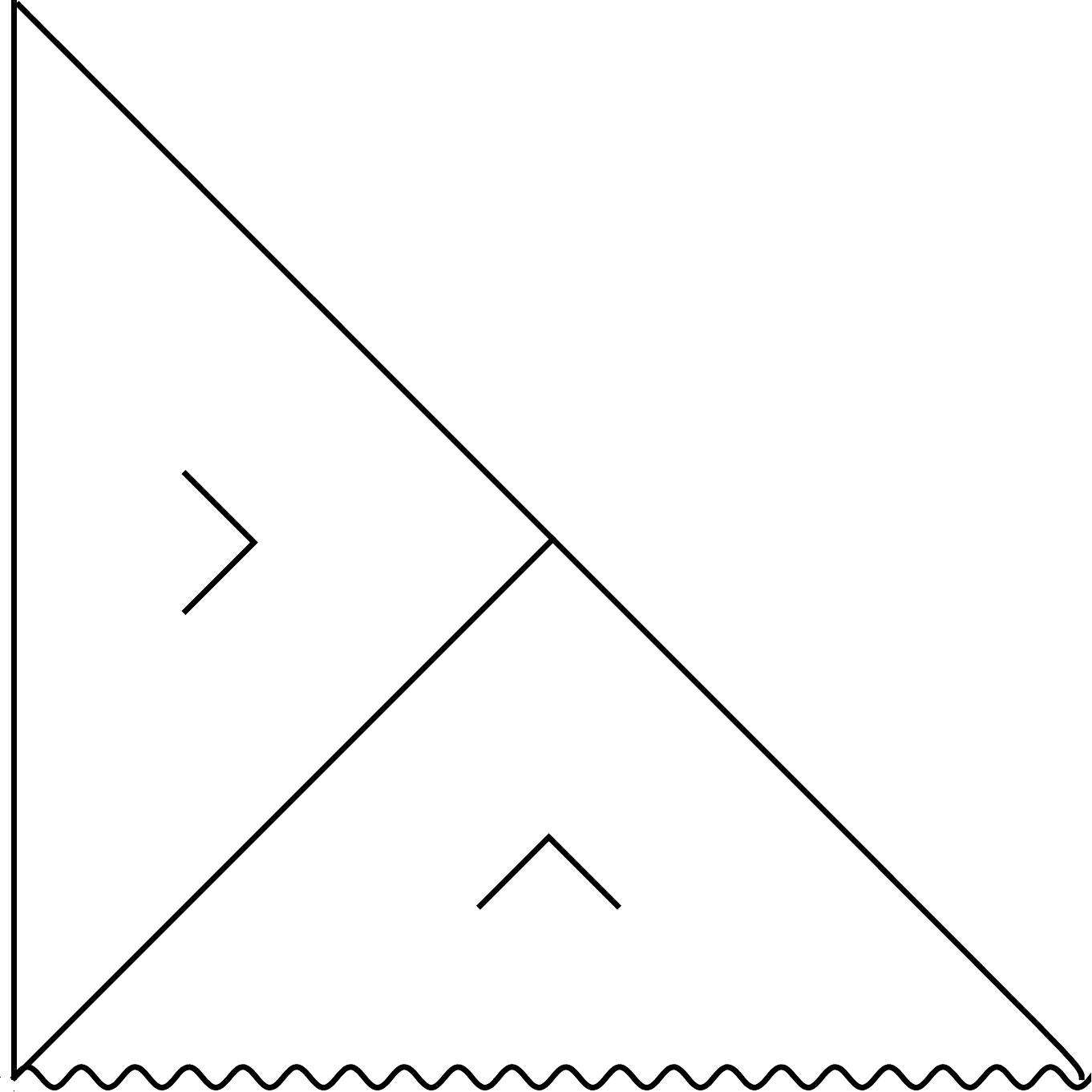}
}
\subfigure[]{
\includegraphics[width=1.5 in]{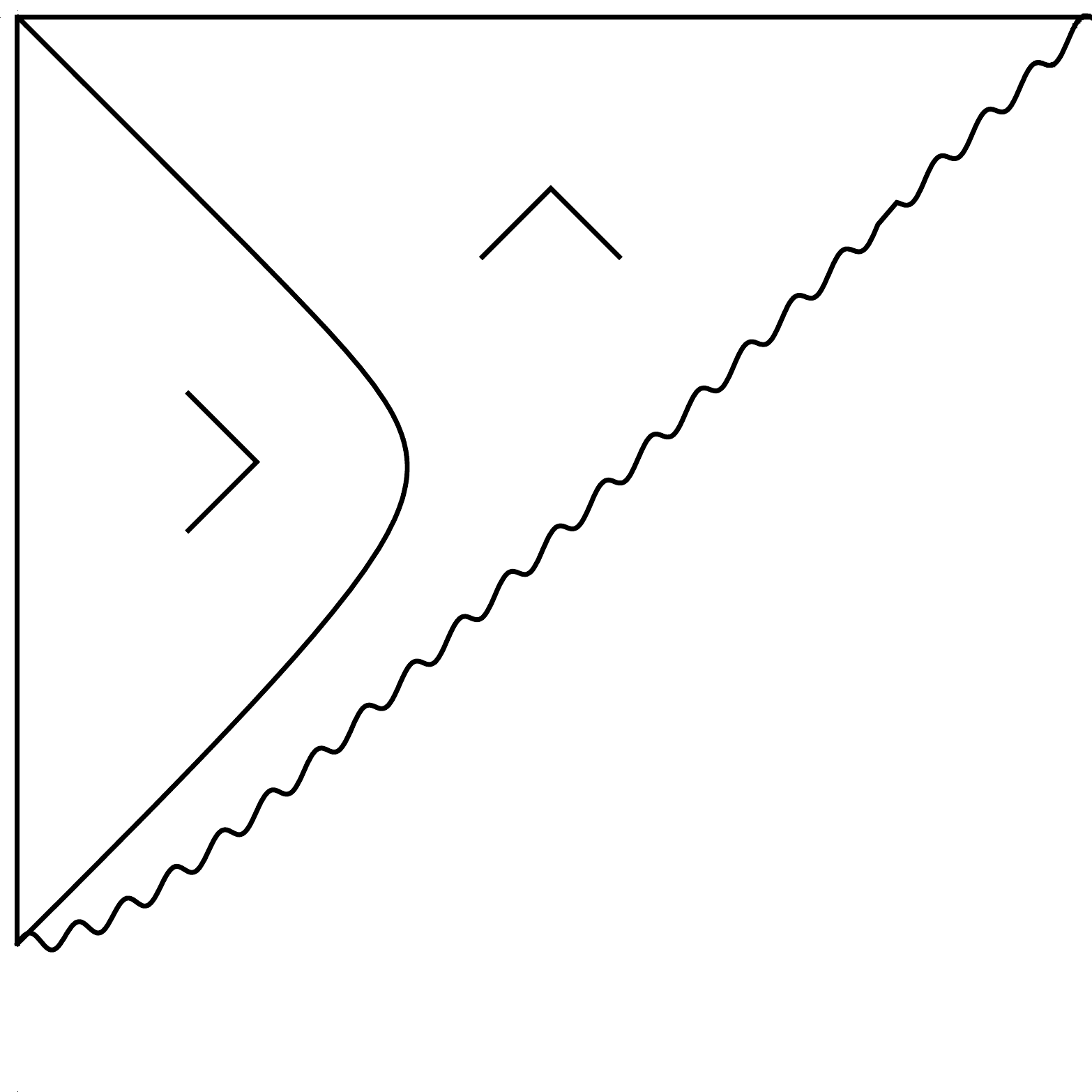}
}
\subfigure[]{
\includegraphics[width=1.5 in]{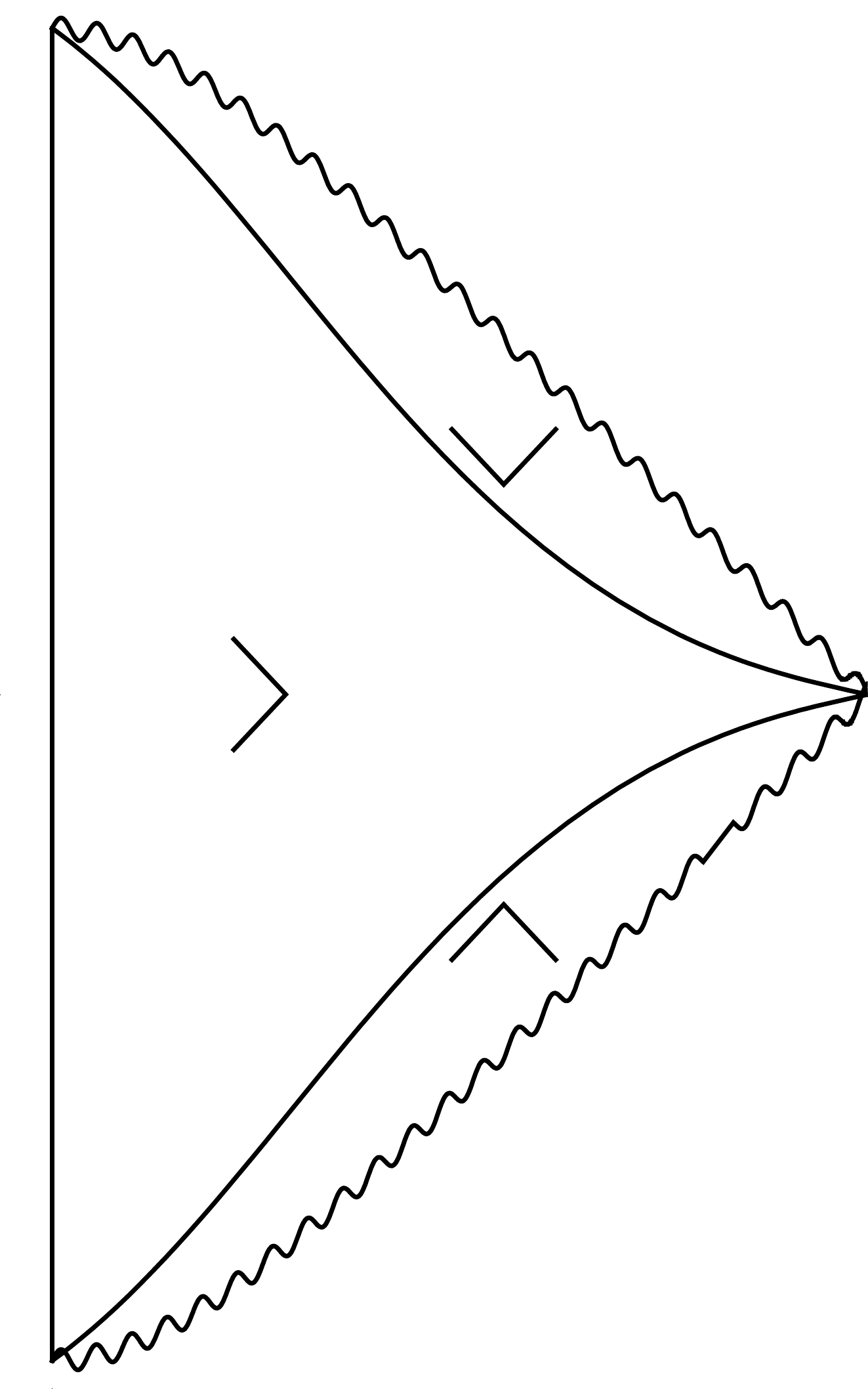}
}
\label{fig-wedges}
\caption{Conformal diagrams for FRW universes with (a) zero, (b) positive, and (c) negative cosmological constant.  Every point represents a sphere, and every sphere has at least two orthogonal null directions along which the area is nonincreasing.  According to the covariant entropy bound, these are the allowed directions along which light-sheets can be constructed.  They are are indicated by wedges. An apparent horizon divides normal ($>$) spheres from trapped ($\vee$) and/or anti-trapped ($\wedge$) spheres.}
\end{figure}

The dynamics is governed by the Friedmann equation:
\begin{equation}
\frac{\dot a^2}{a^2} = \frac{8\pi\rho}{3} + \frac{1}{a^2}~,
\label{eq-fp}
\end{equation}
where $\dot a = da/dt$ and $\rho$ is the total energy density.  
The area of the apparent horizon at the time $t$ is~\cite{RMP}
\begin{equation}
A_{\rm AH}(t)=\frac{3}{2\rho(t)}~.
\label{eq-ah}
\end{equation}
Spheres of smaller area have only ingoing light-sheets (both past and future-directed); the area along outgoing light-rays would increase, violating the nonexpansion condition demanded by the covariant entropy bound.  Spheres outside the apparent horizon posses two past-directed light-sheets, ingoing and outgoing.  In this case, the area increases along future-directed light rays, which are therefore forbidden.   In Fig.~\ref{fig-wedges}, the apparent horizon is shown in a Penrose diagram, and the allowed light-sheet directions are indicated by wedges.

We will consider cosmologies in which the energy density contains two components, radiation\footnote{This is not essential and can be replaced by nonrelativistic matter, as we shall see in Sec.~\ref{sec-matter}.} and vacuum energy, so we may write $\rho=\rho_r+\rho_\Lambda $.  The cosmological constant, $\Lambda=8\pi\rho_\Lambda$, may be positive, negative, or zero.  The energy density of radiation satisfies
\begin{equation}
\frac{8\pi\rho_r}{3} = \frac{t_{\rm c}^2}{a^4} 
\label{eq-rho}
\end{equation}
where the constant $t_{\rm c}$ characterizes the timescale when curvature begins to dominate over radiation.   Up to factors of order unity, the entropy density is given by the $3/4$ power of the radiation energy density:
\begin{equation}
s \sim \rho_r^{3/4} = \frac{t_{\rm c}^{3/2}}{a^3}~. 
\end{equation}

The light-sheets of interest to us originate in, or encompass, the era of curvature domination, when the last term in Eq.~(\ref{eq-fp}) dominates.  With one exception (at the end of Sec.~\ref{sec-nl}), we will not consider cosmologies without such an era, i.e., we will require that
\begin{equation}
t_{\rm c} \ll t_\Lambda ~,
\end{equation}
where $t_\Lambda \equiv |3/\Lambda|^{1/2}$ characterizes the timescale at which vacuum energy begins to dominate the evolution of the universe.
During the curvature era, $t_{\rm c} \ll t\ll t_\Lambda$, Eq.~(\ref{eq-fp}) implies that the scale factor is approximately equal to the age of the universe, $a(t)=t$.
We will find it convenient to use the scale factor, $a$, as a time variable also in other eras, when it does not agree with proper time, $t$.

Now we specify a two-dimensional surface and construct an associated light-sheet.  Consider a sphere of comoving radius $\xi_0$ during the curvature-dominated era, at the time $a_0\gg t_{\rm c}$.  Its proper area is
\begin{equation}
A= 4\pi a_0^2 \sinh^2\xi_0~.
\end{equation}
We demand that the sphere lies on or outside of the apparent horizon, $A\geq A_{\rm AH}$, so that its outgoing past-directed light-sheet exists. That is, the cross-sectional area spanned by outward and past-directed lightrays orthogonal to the initial sphere must be non-expanding.  By setting $ds=0$ in Eq.~(\ref{eq-frwmetric}) and using spherical symmetry, we find that the light-sheet is given by
\begin{equation}
\xi(t) = \xi_0 + \int_t^{t_0} \frac{d\bar t}{a(\bar t)}~.
\end{equation}
Trading $t$ for $a$ as a time variable, this becomes $\xi(a) = \xi_0 + \int_a^{a_0} d\bar{a}/(\bar a \dot{\bar{a}})$.  Using the Friedmann equation,
\begin{equation}
a^2 \dot a^2 = t_{\rm c} ^2+a^2\pm \frac{a^4}{t_\Lambda^2}~,
\label{eq-f2}
\end{equation}
we find that the light-sheet is given by
\begin{equation}
\xi(a) = \xi_0 + \int_a^{a_0} \frac{d\bar{a}}{\sqrt{t_{\rm c}^2 + \bar{a}^2\pm \bar{a}^4/t_\Lambda ^2}}~.
\label{eq-ls}
\end{equation}
The sign in the last term is the sign of $\Lambda$; the term is absent for $\Lambda=0$.

Let $a_1< a_0$ be the time at which the light-sheet ends, and let $\xi_1=\xi(a_1)$.  The entropy on the light-sheet is 
\begin{equation}
S=\sigma V_L ~,
\end{equation}
where 
\begin{equation}
V_L=V_c(\xi_1)-V_c(\xi_0)
\label{eq-vv}
\end{equation}
is the comoving volume occupied by the light-sheet, and
\begin{equation}
\sigma\equiv s a^3\sim t_{\rm c} ^{3/2}
\end{equation}
is the comoving entropy density.

\subsection{Open universes with zero cosmological constant}
\label{sec-zl}
\begin{figure}[tbp]
\centering
   \includegraphics[width=3 in]{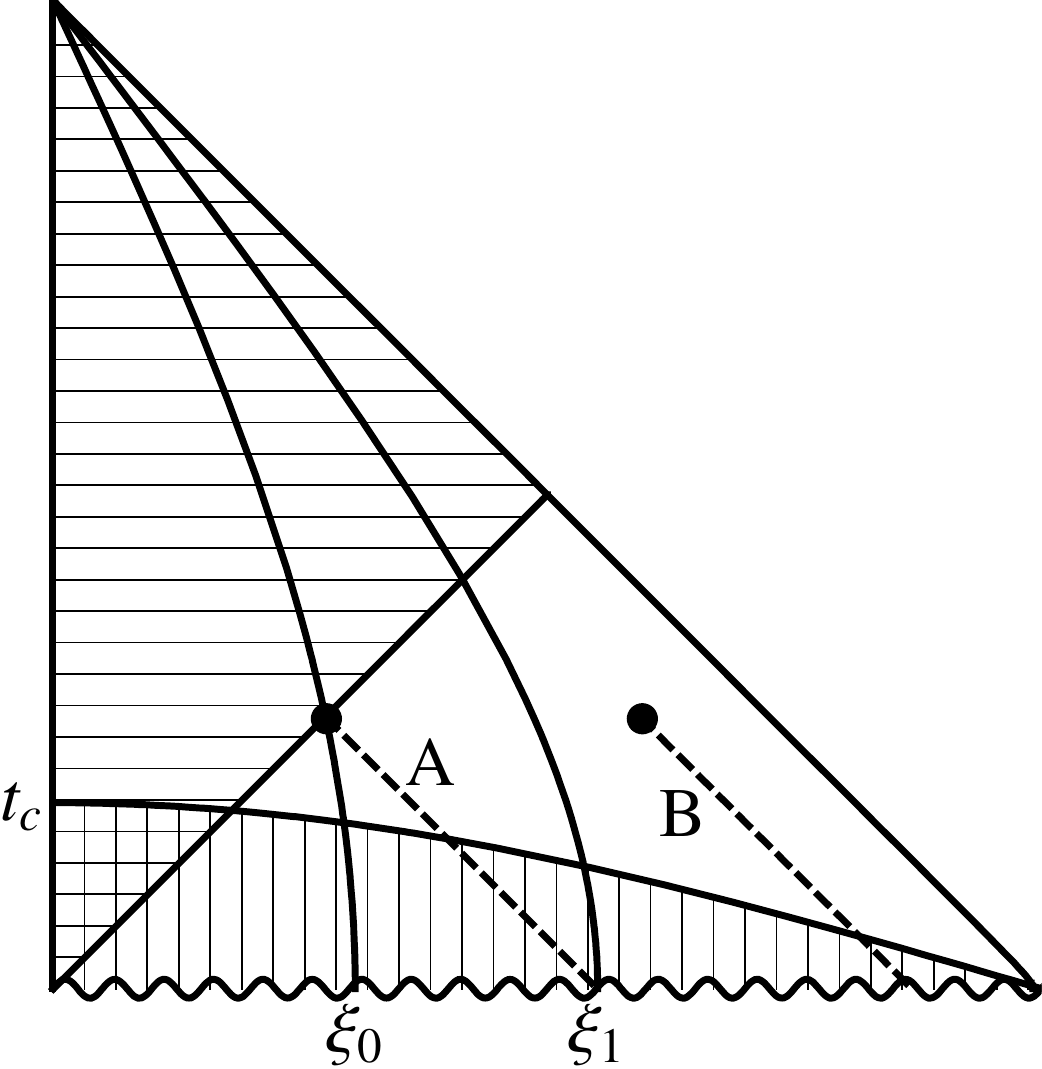}
   \caption{Conformal diagram of an open radiation-dominated $\Lambda=0$ FRW universe.   The spheres of interest to us, whose light-sheets have entropy $S/A\sim t_{\rm c}^{-1/2}$, lie in the unshaded region to the future of $t=t_{\rm c}$ and outside of the apparent horizon.  Two examples are shown.  Light-sheet A starts on the apparent horizon at $\xi=\xi_0$ and ends on the Big Bang at $\xi=\xi_1$.  Light-sheet B starts well outside the apparent horizon.}
   \label{fig-zeroLambda}
\end{figure}

Let us start by setting $\Lambda=0$ (diagram in Fig.~\ref{fig-zeroLambda}); then $\rho=\rho_r$, the last term in Eq.~(\ref{eq-f2}) vanishes, and
Eq.~(\ref{eq-ls}) can be integrated in closed form.  We find that the light-sheet is given by
\begin{equation}
\xi(a) = \xi_0 + \sinh^{-1}(a_0/t_{\rm c})-\sinh^{-1}(a/t_{\rm c})~.
\end{equation}
Since we have chosen $a_0\gg t_{\rm c}$ and $A\geq A_{\rm AH}\sim a^4/t_{\rm c}^2$, we have $\xi_1>\xi_0\gg 1$, so we can approximate $V_c(\xi)\approx (\pi/2)\exp(2\xi)$ in Eq.~(\ref{eq-vv}).  

In the regime $t_{\rm c} \ll a_1\ll a_0$, we thus find $S\sim A t_{\rm c}^{3/2}/a_1^2$ up to factors of order unity.  (Note that $V_c(\xi_0)\ll V_c(\xi_1)$, so make a negligible error by including the entropy missed by the ``hole'' at the center of the light-sheet.) The inverse relation with $a_1$ makes sense: the smaller we chose $a_1$, the longer the light-sheet, and the more entropy it will contain.  

Here we are interested in making $S/A$ as large as possible, so we would like to extend the light-sheet as far as possible. The light-sheet does not encounter caustics and so can be extended all the way to the Big Bang, $a_1\to 0$.  For $a_1\ll t_{\rm c}$, the total entropy on the light-sheet is given by $S\sim \frac{A}{t_{\rm c} ^{1/2}}(1-\frac{2a_1}{t_{\rm c}})$.  Thus, it does not make a big difference whether we terminate the light-sheet at $t_{\rm c}$ or extend it all the way to the Big Bang; up to order-one factors, the entropy is $S\sim A/t_{\rm c}^{1/2}$ for all light-sheets ending at or within the era of radiation domiantion, $a_1\lesssim t_{\rm c} $:
\begin{equation}
\frac{S}{A} \sim t_{\rm c}^{-1/2}~.
\label{eq-sa}
\end{equation}
We have assumed that the sphere $(a_0, \xi_0)$, from which the light-sheet originates, is chosen deep within the regime of curvature domination: $t_{\rm c} \ll a_0$.  This implies $t_{\rm c} ^{-1/2}\gg A^{-1/4}$.  Moreover, the time of curvature domination cannot be earlier than the Planck time: $t_{\rm c} \gtrsim 1$.  From Eq.~(\ref{eq-sa}), we thus obtain
\begin{equation}
A^{3/4}\ll S\lesssim A~.
\end{equation}

We conclude that the entropy in this example exceeds the naive ``bound'' $S\leq A^{3/4}$ by the arbitrarily large ratio $(a_0/t_{\rm c} )^{1/2}$.   Note that $t_{\rm c}$ can be chosen arbitrarily large while keeping this ratio fixed, so the entire light-sheet lies in regions far from the Planck regime.  Thus, the covariant bound and semi-classical gravity are under arbitrarily good control.

We conclude, moreover, the holographic entropy bound can be saturated up to factors of order unity,  $S\sim A$, by choosing the time of curvature domination to be very early, $t_{\rm c} \sim 1$.   In this limit, the initial light-sheet surface, and all but the final edge of the light-sheet, remain in well-controlled regimes where the curvature and densities are small, and the initial area $A$ can be chosen arbitrarily large.  So our example is quite different from the trivial way of saturating the holographic entropy bound ($A\sim 1$, $S\sim 1$), where the whole light-sheet is immersed near or in the Planck regime.  However, most of the entropy is contributed by the outermost rim of the light-sheet, at times of order $t_{\rm c}$, which approaches the Planck time in this limit.  Thus, quantum gravitational corrections start becoming large.  The key point is that the maximum entropy that can be attained on a light-sheet while keeping such corrections small is controlled by the area, not by some smaller power of the area.  By choosing $t_{\rm c}\sim \epsilon^{-2}$, we can achieve $S/A\sim \epsilon$ for arbitrarily large $A$, for any $\epsilon\ll 1$ that controls the size of corrections to semiclassical gravity.

\subsection{Open universes with positive cosmological constant}
\label{sec-pl}

\begin{figure}[tbp]
\centering
   \includegraphics[width=3 in]{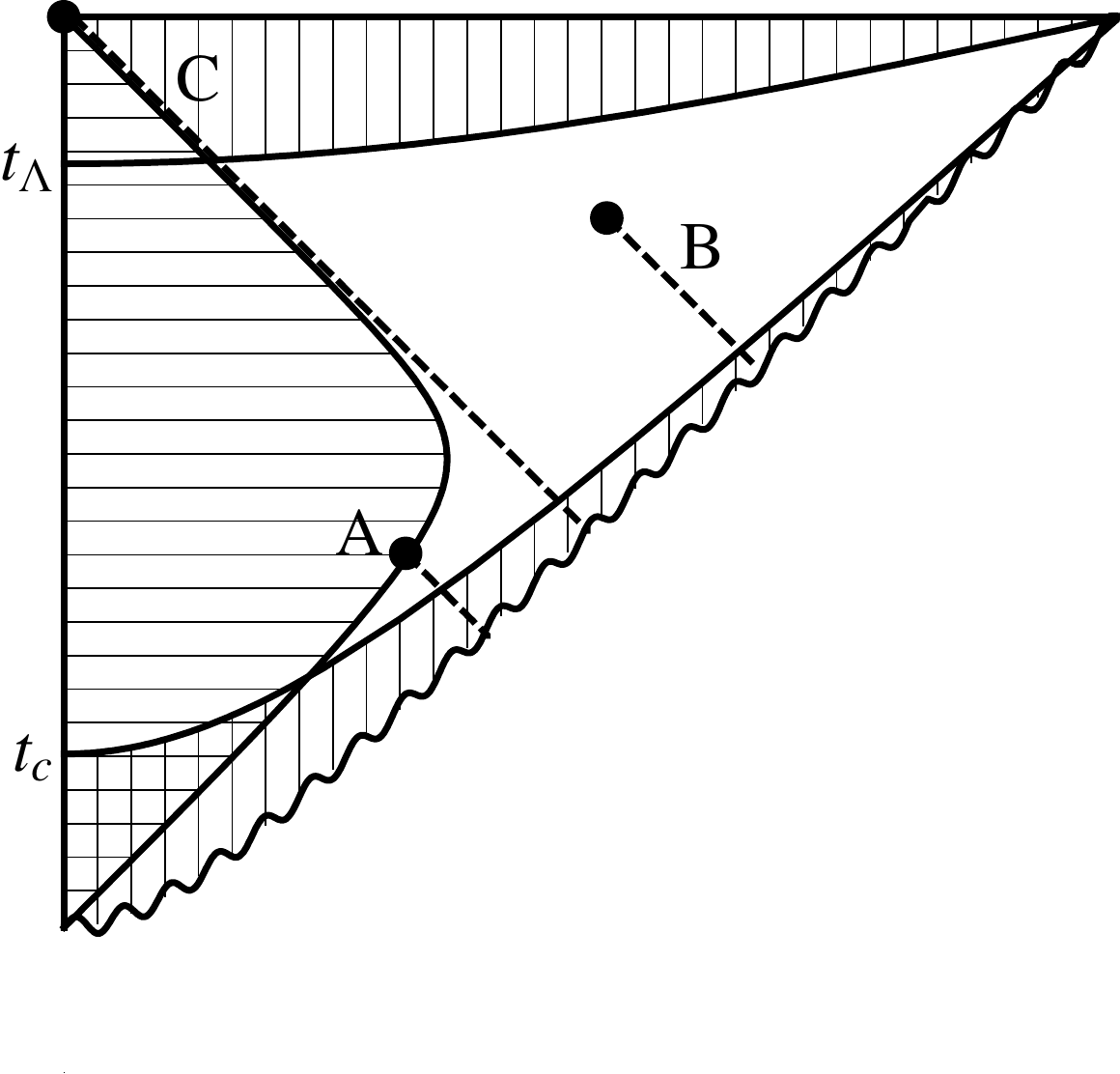}
   \caption{Conformal diagram for an open, radiation-filled FRW universe with $\Lambda>0$.  Again, the light-sheets of interest ($S/A\sim t_{\rm c}^{-1/2}$) lie in the unshaded region.  Light-sheets which start in the region $t>t_{\Lambda}$ have exponentially small $S/A$, except for the light-sheet C, which coincides with the de Sitter event horizon.}
   \label{fig-posLambda} 
\end{figure}

In the presence of a positive cosmological constant, the apparent horizon is given by
\begin{equation}
A_{\rm AH}(t)=\frac{3}{2(\rho_\Lambda+\rho_r)}=
4\pi\left(\frac{t_{\rm c}^2}{a^4}+\frac{1}{t_\Lambda^2}\right)^{-1}~.
\label{eq-ahpos}
\end{equation}
Because of the cosmological term, the apparent horizon area does not grow without bound at late times, but asymptotes to the area of the event horizon of de~Sitter space,
\begin{equation}
A_{\rm dS}=\frac{12\pi}{\Lambda}~,
\end{equation}
for $t\gg t_\Lambda$.  We will again consider only outgoing past-directed light-sheets, which exist for spheres on or outside the apparent horizon, $A\geq A_{\rm AH}$, as shown in Fig.~\ref{fig-posLambda}.

For $t_{\rm c} \ll t_0\lesssim t_\Lambda$ (which implies $a_0\lesssim t_\Lambda$), the $a^4/t_\Lambda^2$ term in Eq.~(\ref{eq-f2}) can be neglected.  Then the analysis is identical to the case $\Lambda=0$.  In particular, the entropy exceeds $A^{3/4}$, and with $t_{\rm c}\to 1$ the bound $S\leq A/4$ can be saturated up to factors of order one.

For $a_0\gg t_\Lambda$, Eq.~(\ref{eq-ls}) becomes
\begin{equation}
\xi_1-\xi_0\sim \int_{t_\Lambda}^{a_0}\frac{t_\Lambda d\bar a}{\bar a^2}
+\int_{t_{\rm c}}^{t_\Lambda} \frac{d\bar a}{\bar a}\sim \log\frac{t_\Lambda}{t_{\rm c}}~,
\end{equation}
where we have set $a_1\sim t_{\rm c}$. (As before, this choice maximizes the entropy for a given initial surface.)  The entropy on this light-sheet is
\begin{equation}
S\sim \sigma V_c(\xi_1)\sim  e^{2\xi_0}\frac{t_\Lambda^2}{t_{\rm c}^{1/2}}~,
\end{equation}
For initial areas containing more than one curvature volume, $\xi_0\gtrsim 1$, the ratio $S/A\sim t_{\rm c}^{-1/2} (t_\Lambda/a_0)^2$ is much smaller than $t_{\rm c}^{-1/2}$, so these light-sheets are of no interest to us.  For $\xi_0\ll 1$, we obtain $S/A\sim t_{\rm c}^{-1/2} (A_{\rm dS}/A)$.  The initial area satisfies $A_{\rm dS}\lesssim A\ll a_0^2$, where the first inequality comes from the requirement $A
\geq A_{\rm AH}$ and the second from the current case, $\xi_0\ll 1$.  If we choose the initial area to be near the lower end of this range, we find again that
\begin{equation}
\frac{S}{A}\sim t_{\rm c}^{-1/2}~.
\label{eq-sads}
\end{equation}
The most interesting special case is obtained by taking the limit $a_0\to\infty$, $\xi_0\to 0$ while holding $A=A_{\rm dS}$ fixed.  The past-directed outgoing light-sheet off of this sphere corresponds to the de~Sitter event horizon.  With early curvature domination, $t_{\rm c}\to 1$, Eq.~(\ref{eq-sads}) implies that the entropy passing through the de~Sitter horizon of a radiation-dominated FRW universe can be comparable to the Bekenstein-Hawking entropy of empty de~Sitter space.  It need not be much less, as has hitherto been believed.  We will discuss this case further in Sec.~\ref{sec-patch}.

\subsection{Universes with negative cosmological constant}
\label{sec-nl}

\begin{figure}[htbp]
\centering
   \includegraphics[width=2.5 in]{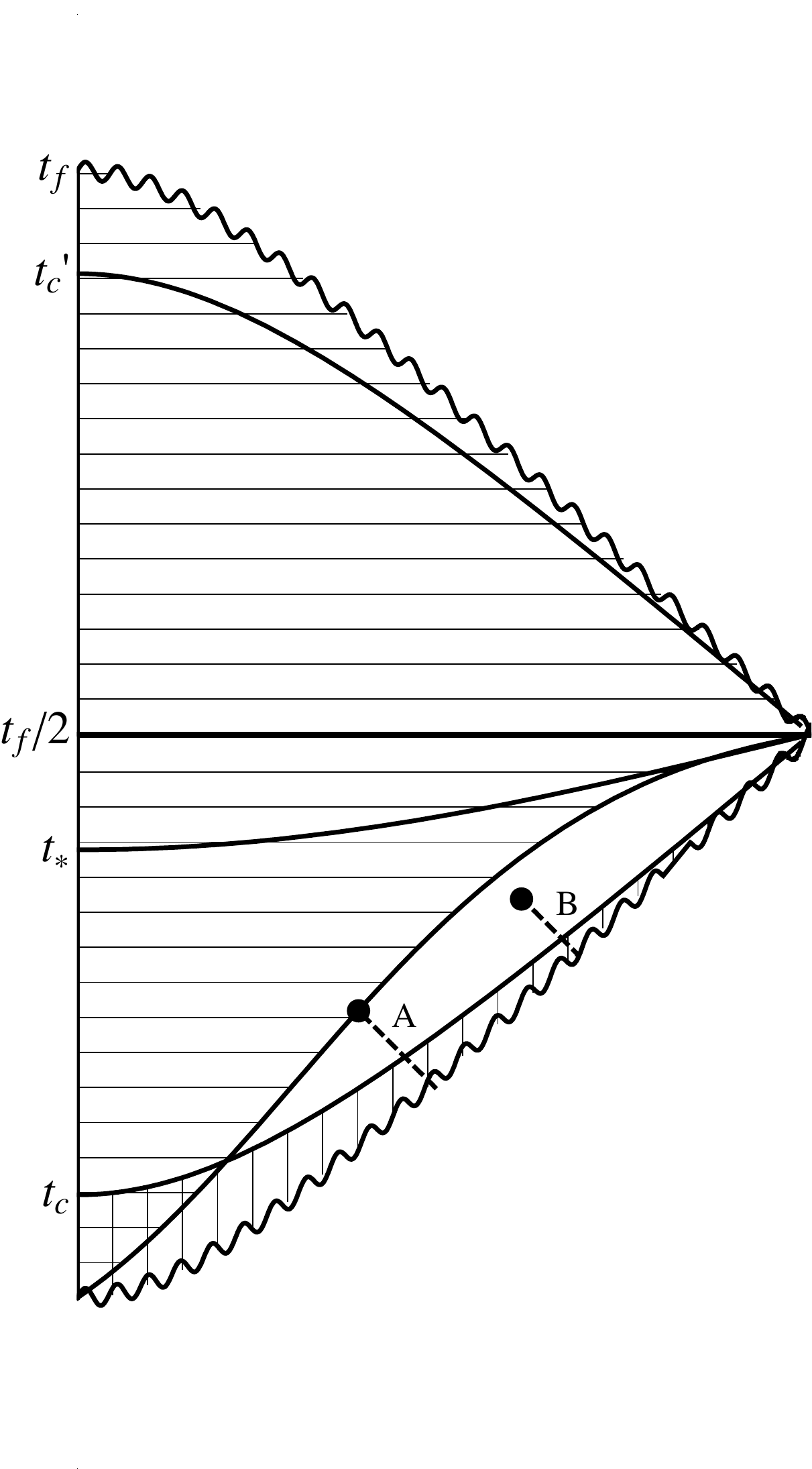}
   \caption{Conformal diagram for an open, radiation-filled FRW universe with $\Lambda<0$, which ends in a Big Crunch. Spheres in the unshaded regions have past-directed outgoing light-sheets with $S/A\sim t_{\rm c}^{-1/2}$.  The apparent horizon diverges at $t_*\equiv (t_\Lambda t_{\rm c})^{1/2}$, which means that there are no anti-trapped spheres with $t>t_*$.}
   \label{fig-negLambda}
\end{figure}

In the presence of a negative cosmological constant (see Fig~\ref{fig-negLambda}), the apparent horizon is given by
\begin{equation}
A_{\rm AH}(t)=\frac{3}{2(\rho_\Lambda+\rho_r)}=
4\pi\left(\frac{t_{\rm c}^2}{a^4}-\frac{1}{t_\Lambda^2}\right)^{-1}~.
\label{eq-ahneg}
\end{equation}
The negative sign of the vacuum energy has important implications.  At the intermediate time
\begin{equation}
t_*=(t_{\rm c} t_\Lambda)^{1/2}~,
\label{eq-tstar}
\end{equation}
the total energy density vanishes and then becomes negative.  
The apparent horizon diverges as $t\to t_*$ and does not exist for $t>t_*$.  Outgoing past-directed light-sheets, which exist for spheres on or outside the apparent horizon, $A\geq A_{\rm AH}(t)$, do not exist at all for $t\geq t_*$, so we require in particular that $a_0<t_*$.  During this entire era, the $a^4/t_\Lambda^2$ term in Eq.~(\ref{eq-f2}) can be neglected, so the light-sheet analysis is identical to the case $\Lambda=0$.  In particular, the entropy exceeds $A^{3/4}$, and with $t_{\rm c}\to 1$ the bound $S\leq A/4$ can be saturated up to factors of order one.

With $\Lambda<0$ there is an additional class of light-sheets with $S\gg A^{3/4}$, which originate from spheres that are normal, i.e., neither trapped nor antitrapped.   These examples have no analogue in universes with nonnegative cosmological constant.  Consider a sphere at $(t_0,\xi_0)$.  We assume that $t_*\leq t_0<t_*'\equiv t_f-t_*$; during this era there is no apparent horizon and all spheres, no matter how large, are normal.  We construct an ingoing light-sheet of this surface, which we may take to be past-directed without loss of generality.  We terminate the light-sheet at the time $t_*$. (Extending it further would complicate our analysis and can only increase the entropy in any case.)  The initial area is of order $a_0^2 \exp(2\xi_0)$.  The comoving radial extent of the light-sheet, $\Delta\xi\equiv \xi_0-\xi(t_*)$, can readily be seen to exceed unity as long as $t_0$ is at least a few times larger than $t_*$.   Therefore its comoving volume is $V_L\approx V_c(\xi_0)$ and the entropy on the light-sheet is of order $e^{2\xi_0} t_{\rm c}^{3/2}$.  We thus find
\begin{equation}
\frac{S}{A}\sim \frac{ t_c^{3/2}}{t_0^2}~.  
\end{equation}
By choosing $t_0\to t_*\equiv (t_{\rm c}  t_\Lambda)^{1/2}$ the ratio $S/A$ can be made as large as $t_{\rm c} ^{-1/2}(t_{\rm c} /t_\Lambda)$ in this region.  Since we have assumed $t_{\rm c}\ll t_\Lambda$, this is smaller than the ratio $t_{\rm c} ^{-1/2}$ attained by antitrapped light-sheets.  Nevertheless, for any value of $t_0$, normal light-sheets in this region violate $S\lesssim A^{3/4}$ by arbitrarily large factors, if we choose the initial sphere sufficiently large.   None of these light-sheets, however, are entirely contained in the past or future light-cone of any event, so we will not revisit them in the next section.  

In a sense, this class of light-sheets continues to exist even for a flat FRW universe ($t_{\rm c}\gg t_\Lambda$) with $\Lambda<0$.  In this case the apparent horizon diverges only at the turnaround time, $t_{\rm f}/2$, when $a\sim t_*$.  This means that at large comoving radius $\xi_0\gg 1$, normal spheres exist only for a very short time and we might as well consider only $t_0=t_{\rm f}/2$.  Terminating the light-sheet at some early time of order $t_\Lambda^2/t_{\rm c}$, one finds that $\Delta\xi\sim t_\Lambda /t_{\rm c} \ll 1$, so the comoving light-sheet volume can be treated as a thin shell.  Therefore, $S\sim t_{\rm c}^{3/2} e^{2\xi_0} \Delta \xi$; and with $A\sim t_*^2 e^{2\xi_0}$, one finds $S/A\sim t_\Lambda^{-1/2}$.   This result connects smoothly with the result in open universes, in the overlap limit $t_{\rm c}\to t_\Lambda$).  However, in the flat FRW case we are forced to extend the light-sheets into the antitrapped region.  Still, it is interesting that we have thus found a cosmological example of $S\gg A^{3/4}$ which is spatially flat and in which the relevant light-sheets originate from normal surfaces.

\subsection{Generalization to pressureless matter}
\label{sec-matter}

In this section, we eliminate the assumption that the universe is filled only with radiation and vacuum energy.  In particular, we will show that light-sheets with entropy $S/A\sim t_{\rm c}^{-1/2}$ exist in an open FRW universe filled with pressureless dust.  We will set $\Lambda=0$ in this section, though our results do generalize straightforwardly to universes with nonzero cosmological constant, as long as $t_\Lambda \gg t_{\rm c}$, just as they did for radiation.

The energy density of pressureless dust satisfies
\begin{equation}
\frac{8\pi\rho}{3}=\frac{t_{\rm c}}{a^3}~.
\end{equation}
With $\Lambda=0$, the Friedmann equation reads
\begin{equation}
a^2\dot a^2 = t_{\rm c} a + a^2~.
\label{eq-nonrelfried}
\end{equation}
As before, we consider a sphere of radius $\xi_0 > \xi_{AH} \approx \frac{1}{2}\log a_0/t_{\rm c}$, and area $A\sim a_0^2 e^{2\xi_0}$, during the curvature-dominated era, $a_0 \gg t_{\rm c}$.  Its outward-going, past-directed light-sheet, parametrized by $a$, has radius
\begin{align}
\xi(a) &= \xi_0 + \int_a^{a_0} \frac{d\bar{a}}{\sqrt{t_{\rm c}\bar{a} + \bar{a}^2}} \\
&= \xi_0 + 2\sinh^{-1}\left(\sqrt{a_0/t_{\rm c}}\right)- 2\sinh^{-1}\left(\sqrt{a/t_{\rm c}}\right)~.\label{eq-ximatter}
\end{align}
The entropy density is of order the number density of particles:
\begin{equation}
s \sim \frac{\rho}{m} \sim \frac{t_{\rm c}}{ m a^3}~,
\label{eq-nonrel}
\end{equation}
where $m$ is the particle mass; the comoving entropy density is therefore $\sigma\sim t_{\rm c}/m$.   We extend the light-sheet as far as possible to maximize the entropy on it: $a_1\to 0$.  As in Sec.~\ref{sec-main}, however, the dominant contribution comes from the curvature dominated era, so we may as well set $a_1\sim t_{\rm c}$ in Eq.~(\ref{eq-ximatter}).   The comoving volume covered by this light-sheet is $V_L\sim e^{2\xi_1}\sim e^{2\xi_0} a_0^2/t_{\rm c}^2$.  With $S=\sigma V_c$, we find
\begin{equation}
\frac{S}{A} \sim (mt_{\rm c})^{-1}~.
\label{eq-samatter}
\end{equation}
For $m t_{\rm c}<A^{1/4}$, the entropy exceeds the naive bound $S\lesssim A^{3/4}$.   As $m t_{\rm c} \to 1$, the holographic bound can be saturated, $S/A\to 1$.   So far, so good: this confirms our claim that high-entropy light-sheets can be found not only in regions with radiation but also with nonrelativistic particles.  What is important is the open spatial geometry of the FRW solution, not the matter content.  

But Eq.~(\ref{eq-samatter}) should make us suspicious: What is to prevent us from choosing $m t_{\rm c} \ll 1$ and violating the holographic bound?  Our formula for the entropy density, Eq.~(\ref{eq-nonrel}), is valid only if the dust particles are dilute, i.e., nonoverlapping.  This is the case if the number density of particles, $\rho/m$, is less than one particle per Compton wavelength cubed, i.e., if
\begin{equation}
m^4>\frac{t_{\rm c}}{a^3}~.
\end{equation}
If we want to use Eq.~(\ref{eq-nonrel}) along the whole light-sheet, we must ensure that the above condition holds for all $a$ in the interval $a_1\leq a\leq a_0$.  Thus, we require
\begin{equation}
m>t_{\rm c}^{-1/2}~.
\end{equation}
If this inequality is saturated, i.e., if the matter first becomes dilute around the time $t_{\rm c}$), we recover the relation $S/A \sim t_{\rm c}^{-1/2}$ once more.  If it is violated, the entropy density will be smaller than that of Eq.~(\ref{eq-nonrel}) and the light-sheet will contain less entropy than that given in Eq.~(\ref{eq-samatter}).  Thus, the holographic bound cannot be violated by pressureless matter, though it can be saturated.

\section{Maximizing observable entropy}
\label{sec-patch}

In the previous section, we found classes of light-sheets whose entropy is large compared to naive expectations ($S\gg A^{3/4}$), and which can even saturate the covariant bound ($S\sim A$).  In this section, we show that a subclass of these light-sheets is entirely contained in the past of a single event.  In this sense, both the light-sheets and the corresponding entropy are observable. Moreover, we will show that the maximum observable radiation entropy is of order $\Lambda^{-1}$ if the cosmological constant is positive, and $\Lambda^{-2}$ if it is negative.

We consider the same FRW cosmologies as before: open; filled with radiation; with positive, negative, or zero cosmological constant; and with a long curvature-dominated era ($t_{\rm c} \ll t_\Lambda $).   By homogeneity, we need only consider an observer at the origin, $\xi=0$.   At the time time $t_E$, let $L(t_E)$ be the {\em past light-cone\/} of the observer; i.e., $L(t_E)$ is the boundary of the past of an event at $t=t_E$, $\xi=0$.  $L(t_E)$ can be obtained from Eq.~(\ref{eq-ls}) by setting $\xi_0\to 0$ and $a_0\to a(t_E)$.  The observer's {\em causal patch\/} is the (disjoint) union of all these past light-cones; this is the spacetime region that the observer can ever receive signals from.   (If $\Lambda>0$, this is the interior of the event horizon; if $\Lambda=0$, it is the entire spacetime.)

\subsection{Entropy on the past light-cone at finite time}
\label{sec-cone}

\begin{figure}[htbp]
\centering
   \includegraphics[width=3 in]{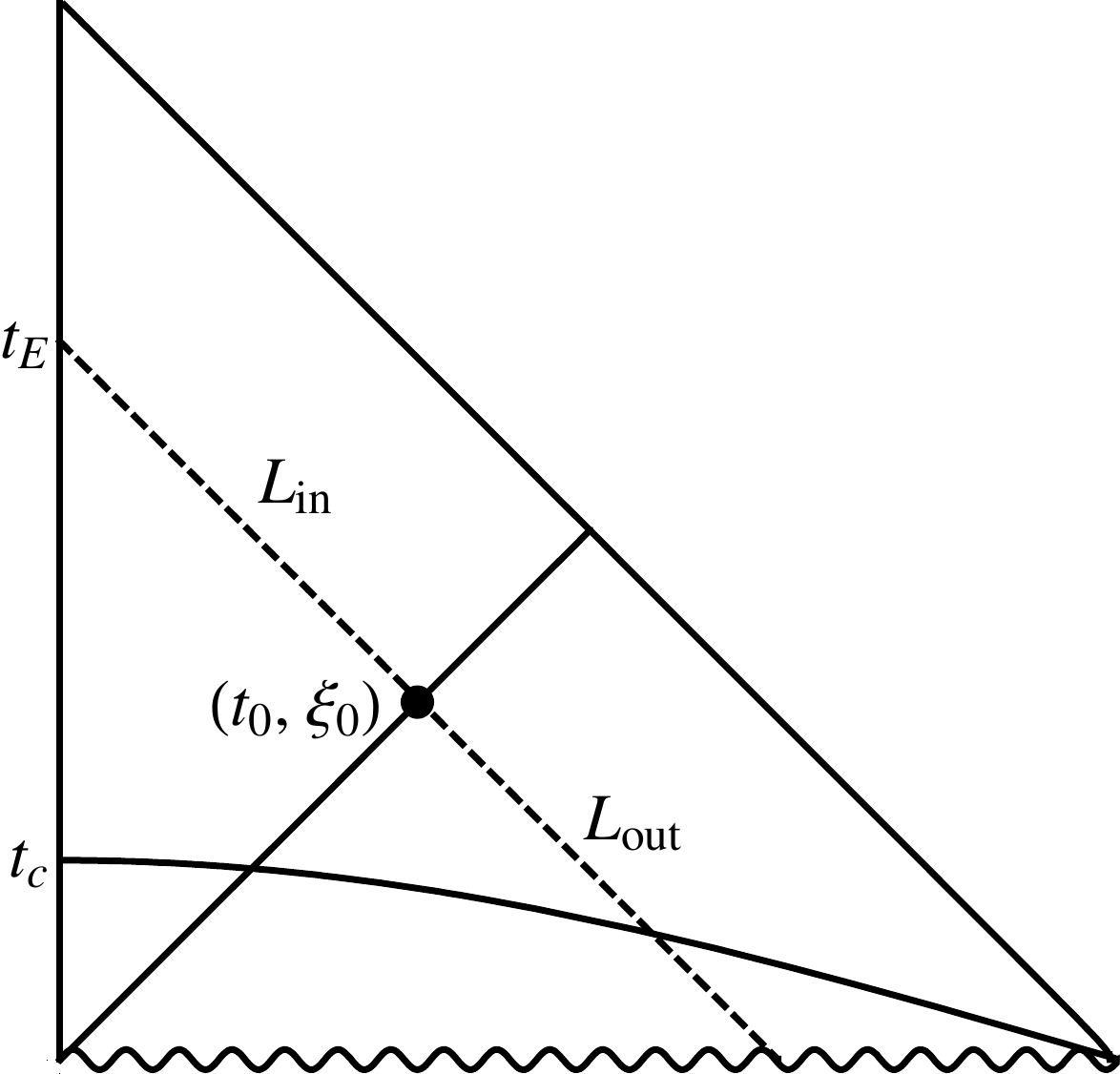}
   \caption{The dashed line shows the past light-cone of an observer at $\xi=0$ at the time $t_E\gg t_{\rm c}$, in a $\Lambda=0$ radiation-dominated open FRW universe.     Its area is maximal on the apparent horizon, at $(t_0,\xi_0)$.  This sphere has two light-sheets, $L_{\rm in}$ and $L_{\rm out}$, which together comprise the past light-cone.  The entropy on $L_{\rm out}$ satisfies $S/A\sim t_{\rm c}^{-1/2}$; the entropy on $L_{\rm in}$ is negligible by comparison.  The total entropy in the observer's past is thus of order $A t_{\rm c} ^{-1/2}$, where $A$ is the largest observed sphere. }
   \label{fig-ah}
\end{figure}

To demonstrate that a light-sheet constructed in the previous section is contained in an observer's past, it suffices to show that it lies on one of the past light-cones of the observer.  The cross-sectional area of any past light-cone $L(t_E)$ in a Big Bang cosmology vanishes both at its apex at $t_E$, and at the Big Bang.  Therefore,  every past light-cone possesses a sphere of maximum area, or apparent horizon.  Let  $t_0$ be the time when this maximum occurs.  It can be shown that $t_0$ is a monotonically growing function of $t_E$ in the cosmologies considered here. 

The apparent horizon divides $L(t_E)$ into two pieces (see Fig.~\ref{fig-ah}): the ``inner'' piece, $L_{\rm in}$, covers the range $t_0\leq t\leq t_E$ from the apparent horizon up to the apex $t<t_0$; the ``outer'' piece, $L_{\rm out}$,  ranges from the apparent horizon down to the Big Bang ($t_0\geq t> 0$).  Because the apparent horizon is a local maximum, the cross-sectional area  is decreasing away from the apparent horizon along both pieces.  In other words, the apparent horizon is a surface from which two light-sheets $L_{\rm in}$ and $L_{\rm out}$ originate in opposite directions, which together make up the past light-cone $L$~\cite{CEB2}.  The entropy in the past of an observer at the time $t_E$ is the sum of the entropy on these two light-sheets.  The light-sheet $L_{\rm in}$ is future-directed and ingoing.  The light-sheet $L_{\rm out}$ is past-directed and outgoing, like the light-sheets we considered in the previous section.   Thus, it {\em may\/} contain large entropy, in the sense that 
\begin{equation}
\frac{S(L_{\rm out})}{ A_{\rm AH}[t_0(t_E)]}~ t_{\rm c}^{-1/2}~.
\label{eq-sacone}
\end{equation}
We will now show that it {\em does}, provided that $t_E\gg t_{\rm c}$.



For $\Lambda=0$, we showed in Sec.~\ref{sec-main} that the entropy on a past-directed outgoing light-sheet will satisfy $S/A\sim t_{\rm c}^{-1/2}$
if and only if its initial surface is well inside the regime of curvature domination, $t_0\gg t_{\rm c}$.  For the light-sheet $L_{\rm out}$ that constitutes the ``outer'' portion of a past light-cone with apex at $t_E$, the condition $t_0\gg t_{\rm c}$ is satisfied if and only if the apex itself occurs well inside this regime, i.e., if $t_E\gg t_{\rm c}$.

To prove this, assume that $t_0\gg t_{\rm c}$ and set $a\to t_E$ and $\xi(a)\to 0$ in Eq.~(\ref{eq-ls}).   Since the integration takes place entirely in the curvature-dominated era and we have assumed $\Lambda=0$, we can neglect the first and last term under the square root.  By Eqs.~(\ref{eq-ah}) and (\ref{eq-rho}), the apparent horizon is located at $\xi_0\sim \log (t_0/t_{\rm c})$.  Solving for $t_E$, we find
\begin{equation}
t_E\sim \frac{t_0^2}{t_c}~.
\end{equation}
It follows that the conditions $t_E\gg t_{\rm c}$ and $t_0\gg t_{\rm c}$ are equivalent, as claimed.  

In the presence of a cosmological constant, $\Lambda\neq 0$, Eq.~(\ref{eq-sacone}) continues to hold, even for the past light-cones of events in the vacuum-dominated era, $t_E\gtrsim t_\Lambda$.  This follows from the monotonicity of $t_0(t_E)$ and from our result that all past-directed outgoing light-sheets with $t_0\gg t_{\rm c}$ (for $\Lambda<0$), or with $t_0\gg t_{\rm c}$ and $A\lesssim A_{\rm dS}$ (for $\Lambda>0$), satisfy $S/A\sim t_{\rm c} ^{-1/2}$.  Note that the case discussed before Eq.~(\ref{eq-sads}), in which the initial area is much larger than $A_{\rm dS}$ and $S/A\sim t_{\rm c} ^{-1/2}$ does not hold, cannot arise for any light-sheet that lies on the past light-cone of an event, since $A_{\rm AH}<A_{\rm dS}$ by Eq.~(\ref{eq-ahpos}.) 

We conclude that for all times $t_E\gg t_{\rm c}$, an observer's causal past contains light-sheets that satisfy Eq.~(\ref{eq-sa}).  In this era, the entropy in the causal past of an observer is
\begin{equation}
S(L)\sim A_{\rm AH}~ t_{\rm c}^{-1/2}~,
\label{eq-sacone2}
\end{equation}
where $A_{\rm AH}$ is the maximum area on the boundary of the observer's past, and thus, the largest sphere in the observer's past.  [Note that $S(L)=S(L_{\rm in})+S(L_{\rm out})$ but the contribution from the inner light-sheet is negligible.]  The observable entropy exceeds the naive bound $A_{\rm AH}^{3/4}$ (which does apply in flat FRW universes), by the arbitrarily large factor $t_E^{1/2}$ for $\Lambda\leq 0$, or by $\min\{t_E,t_\Lambda\}^{1/2}$ for $\Lambda>0$.  It can be comparable to the apparent horizon area, $S(L)/A_{\rm AH}\to 1$, in the limit of early curvature domination, $t_{\rm c}\to 1$.  

\subsection{Entropy in the causal patch}
\label{sec-patchsub}

Next, let us ask what the largest amount of entropy is that can ever be observed, i.e., the entropy in the causal patch, in a universe with cosmological constant $\Lambda\neq 0$.  (With $\Lambda=0$, the observable entropy is unbounded in the limit $t_E\to \infty$.)  In a flat radiation-dominated FRW universe, the maximum observable entropy is $|\Lambda|^{-3/4}$; this has sometimes been treated as a universal bound~\cite{LinVan09}.  We will now show that the observable entropy is much larger in an open universe:  $S_{\rm max} \sim \Lambda^{-1}$ for $\Lambda>0$ and $S_{\rm max} \sim \Lambda^{-2}$ for $\Lambda<0$.

\paragraph{Positive cosmological constant}
For $\Lambda>0$, we are interested in the limit as $t_E\to \infty$, in which the past light-cone $L(t_E)$ becomes the de~Sitter event horizon.  This null hypersurface is special: it is not the boundary of the past of any event, and its cross-sectional area is everywhere decreasing towards the past.  Thus, it consists not of two light-sheets, but only of a single light-sheet, with initial area given exactly by $A_{\rm dS}$. In Sec.~\ref{sec-main}, we obtained the same light-sheet by taking a slightly different limit after Eq.~(\ref{eq-sads}).  All radiation that is observable in principle passes through the event horizon, so $S(L)=S_{\rm max}$ for this light-sheet, and Eq.~(\ref{eq-sads}) implies
\begin{equation}
S_{\rm max}\sim \frac{A_{\rm dS}}{t_{\rm c}^{1/2}}\sim \Lambda^{-1}\, t_{\rm c}^{-1/2}~.
\end{equation}
For $t_\Lambda \gg t_{\rm c}$, this vastly exceeds the naive bound $S_{\rm max}< \Lambda^{-3/4}$.  In the limit of early curvature domination, the radiation entropy passing through the event horizon of an asymptotically de~Sitter, radiation-filled universe can be comparable to the final entropy, the Bekenstein-Hawking entropy of the de~Sitter horizon:
\begin{equation}
S_{\rm max}\to \Lambda^{-1} ~~\mbox{for}~~t_{\rm c} \to 1~.
\end{equation}

\paragraph{Negative cosmological constant}
For $\Lambda<0$, the evolution of the universe is symmetric about the time of maximum expansion, $t_{\rm turn}\sim a_{\rm max}\approx t_\Lambda$.  In the limit as $t_E$ approaches the Big Crunch, the observer's past light-cone extends out to $\xi_{\rm max}=2\xi_{\rm turn}$, where
\begin{equation}
\xi_{\rm turn}\approx \int_0^{a_{\rm max}} \frac{d\bar{a}}{\sqrt{t_{\rm c}^2 + \bar{a}^2+\bar{a}^4/t_\Lambda ^2}}\approx
\int_{t_{\rm c}}^{a_{\rm max}} \frac{d\bar{a}}{\bar{a}}\approx \log \frac{t_\Lambda}{t_{\rm c}}
\end{equation}
is the comoving distance travelled by a light-ray during each half of the universe's history.  The entropy within a comoving sphere of radius $\xi$ is $S(\xi)=\sigma V_c(\xi)$.  Using $\sigma\sim t_{\rm c}^{3/2}$ and $V_c\sim e^{2\xi_{\rm max}}=e^{4\xi_{\rm turn}}$, we find that the entropy in the causal patch is
\begin{equation}
S_{\rm max} \sim \frac{t_\Lambda ^4}{t_{\rm c}^{5/2}}\sim \Lambda^{-2}\,t_{\rm c}^{-5/2}~.
\end{equation}
This shows that the observable entropy in an open universe with negative cosmological constant can vastly exceed the naive bound $|\Lambda|^{-3/4}$ (which does hold in spatially flat FRW universes.   In the limit of early curvature domination, the observable entropy approaches
\begin{equation}
S_{\rm max}\to \Lambda^{-2} ~~\mbox{for}~~t_{\rm c} \to 1~.
\end{equation}

For the record we note some properties of the maximum area of the past light-cone in the limit where $t_E$ approaches the crunch.  From the monotonicity of $t_0(t_E)$ and Eq.~(\ref{eq-tstar}) we know that the boundary of the causal patch intersects the apparent horizon long after curvature domination but long before the turnaround: $t_{\rm c} \ll t_0\ll t_\Lambda$.   At this time, the comoving radius of the boundary of the patch is $\xi(t_0)=\xi_{\rm max}-\log{t_0}{t_{\rm c}}$, and its proper area is of order 
\begin{equation}
A_{\rm AH}\sim t_0^2 e^{2\xi(t_0)}\sim \Lambda^{-2} t_{\rm c} ^{-2}~.
\end{equation}
[Comparison wit Eq.~(\ref{eq-ahneg}) reveals that $t_0\approx t_*$ but we will not need this result here.]  Note that $S_{\rm max}/A_{\rm AH}\sim t_{\rm c}^{-1/2}$.  This is consistent with Eq.~(\ref{eq-sa}), since $S_{\rm max} \approx  S(L_{\rm out})$, i.e., the entropy on the future-directed ingoing light-sheet of the apparent horizon sphere is negligible compared to the entropy on the past-directed outgoing light-sheet, to which Eq.~(\ref{eq-sa}) applies.  Note also that the area of the largest sphere in the causal patch can thus be much greater than $|\Lambda|^{-3/4}$, its value in spatially flat FRW universes with negative cosmological constant.   In the limit $t_{\rm c}\to 1$ the maximum area approaches $\Lambda^{-2}$.

\section{Saturating the entropy bound in gravitational collapse}
\label{sec-ball}

The cosmological solutions we have discussed thus far are interesting because they lead to situations where the entropy is maximized in the past light-cone of an observer (as discussed in Sec.~\ref{sec-patch}). The initial conditions for these solutions are specified on a large spatial slice near the Big Bang, consisting of many causally disconnected regions. Such a slice is not in the future light-cone of any observer, and so it would be impossible for an experimentalist to create such a system from scratch. In this section we will explore solutions which describe systems an experimentalist could create starting in flat space, and which also contain a light-sheet that beats the naive $S\sim A^{3/4}$ bound. The key difference between the present discussion and that of Sec.~\ref{sec-patch} is that here we want the entropy-saturating light-sheet to be contained in the future light-cone of a single observer, rather than the past light-cone.

\begin{figure}[tbp]
\centering
   \includegraphics[width=3 in]{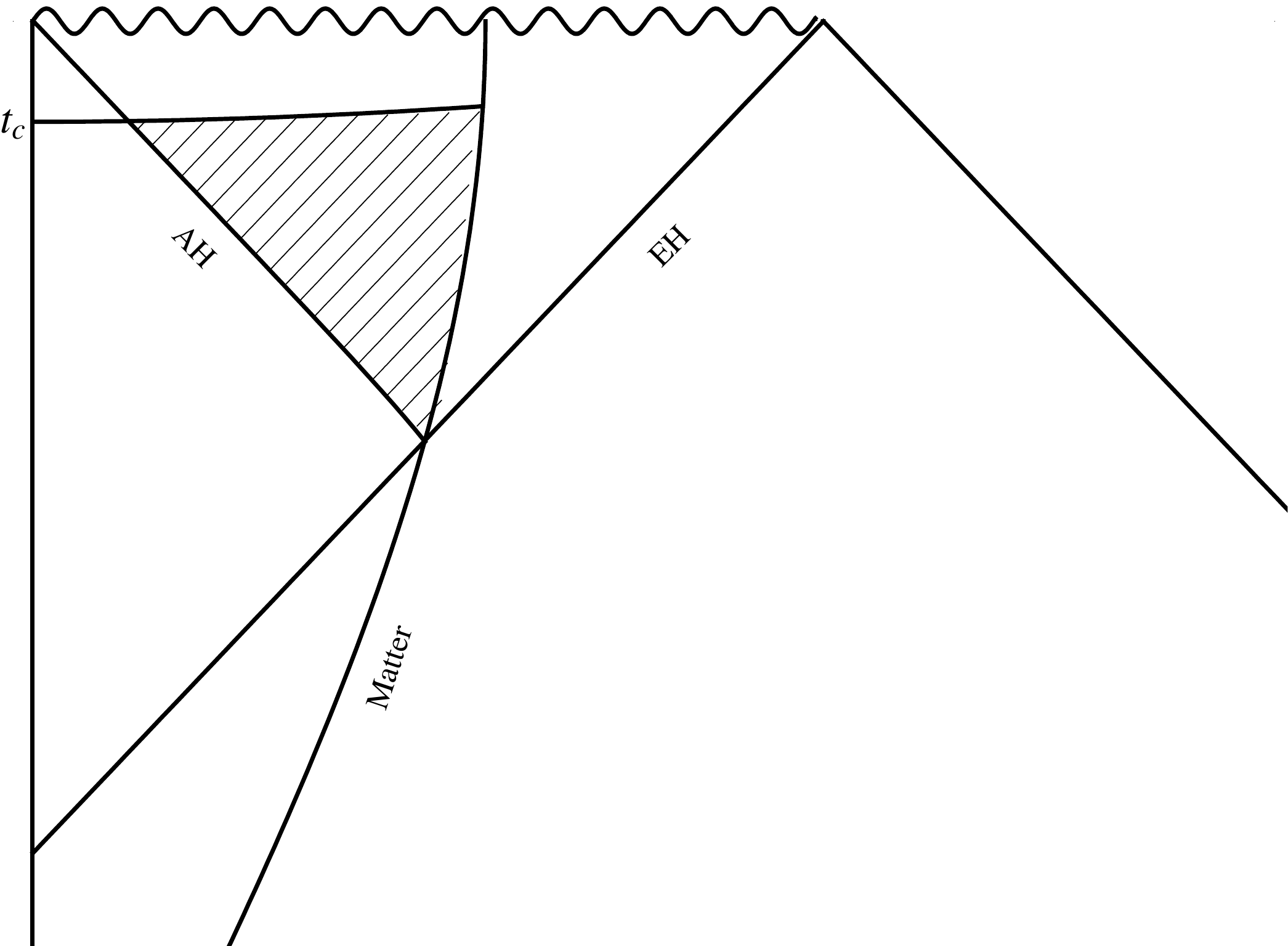}
   \caption{Conformal diagram for a collapsing ball of pressureless matter, the edge of which is labeled ``Matter''.  We consider the future-directed outgoing light-sheets of spheres in the shaded region, deep inside the black hole event horizon. Light-sheets which remain inside the matter ball until $t_{\rm c}$ satisfy $S/A\sim t_{\rm c}^{-1/2}$.  (Light-sheets which reach the edge of the matter ball before $t_{\rm c}$ satisfy $S/\Delta A\sim t_{\rm c}^{-1/2}$; see Appendix \protect\ref{sec-geb}.)}
   \label{fig-bh}
\end{figure}

\subsection{A collapsing ball of dust}

The solution for pressureless matter we presented in Sec.~\ref{sec-matter} approaches flat Minkowski space in the far future.
By time-reversing the solution, we find a collapsing universe which is arbitrarily dilute, and arbitrarily flat, in the far past.
The observer of Sec.~\ref{sec-patch}, whose {\em past} light-cone in the original solution contained the saturating light-sheet, becomes an observer whose {\em future} light-cone contains the saturating light-sheet.
The time-reversed cosmology will be our model for a collapsing ball of pressureless dust created by an experimentalist. 
The dust ball consists of the time-reversed cosmology out to a maximal comoving radius $\xi_{\rm max}$, together with vacuum outside.
By Birkhoff's theorem, the geometry outside of the ball is just a portion of the Schwarzschild black hole metric:
 \begin{equation}
ds^2 = -\left(1-\frac{2M}{r}\right)d\tau^2 + \left(1-\frac{2M}{r}\right)^{-1}dr^2 + r^2 d\Omega^2~.
\end{equation}
Our first task is to determine the relationship between the Schwarzschild variables $(M, r, \tau)$ describing the space outside of the ball, and the cosmological variables $(t_{\rm c} , \xi, t)$ describing the dust ball itself.

Consider a geodesic in the cosmology at fixed comoving coordinate $\xi_{\rm max}$, which is the boundary between the FRW and Schwarzschild regions.
It is the worldline of one of the dust particles on the edge of the dust ball.
As such it must be a radial, timelike geodesic from both the FRW point of view and the Schwarzschild point of view.
The FRW coordinate time $t$ is the proper time along the geodesic, so its relation to Schwarzschild coordinates is
\begin{equation}
dt^2 = (1-2M/r)d\tau^2 - (1-2M/r)^{-1}dr^2~.
\end{equation}
Schwarzschild geometry possesses a timelike Killing vector $\tau^a$, and by contracting with the four velocity $v^a$ of the geodesic we can construct a dimensionless conserved quantity, $\gamma$:
\begin{equation}
\gamma = -g_{ab}\tau^a v^b = \left(1-\frac{2M}{r}\right) \frac{d\tau}{dt}~.
\end{equation}
This equation is more useful in the form
\begin{equation}
\frac{\dot{r}^2}{r^2} =  \frac{2M}{r^3}+\frac{\gamma^2-1}{r^2}~,
\end{equation}
where $\dot{r} = dr/dt$.
At some  fixed $t$, the collection of all comoving geodesics at radius $\xi_{\rm max}$ form a $2$-sphere of area $a(t)^2\sinh^2\xi_{\rm max}$.
This must match the area obtained for the same surface using the Schwarzschild metric, which gives us the relation
\begin{equation}
r(t) = a(t)\sinh \xi_{\rm max}~.
\end{equation}
Upon substitution, the conservation equation for $\gamma$ becomes Eq.~(\ref{eq-nonrelfried}), the Friedmann equation:
\begin{equation}
\frac{\dot{a}^2}{a^2} =  \frac{2M}{a^3 \sinh^3\xi_{\rm max}}+\frac{\gamma^2-1}{a^2\sinh^2 \xi_{\rm max}}~.
\end{equation}
By inspection we discover that $\gamma = \cosh \xi_{\rm max}$ and $2M = t_{\rm c} \sinh^3 \xi_{\rm max}$.

Further insight is gained by examining the dust ball in the dilute limit.
In the dilute limit (that is, $t\rightarrow \infty$ in FRW coordinates and $r\rightarrow \infty$ in Schwarzschild coordinates), both metrics reduce to Minkowski space and we can ask how the matter is distributed on a constant-Minkowski-time slice.
The Schwarzschild coordinates $(\tau,\,r)$ approach the usual time and radial coordinates of Minkowski space, while the FRW coordinates $(t, \, \xi)$ reduce to Milne coordinates:
\begin{align}
\tau&= t\cosh \xi,&r=t\sinh \xi~.
\end{align}
On a slice of constant $\tau= \tau_0$, the particles of matter occupy (be definition) the region of space $0\leq r \leq r_{\rm max}$ and move in the radial direction.
Since the comoving geodesics are paths of constant $\xi$, the particle speeds on constant-$\tau$ slices are given by $v= dr/d\tau= \tanh \xi$.
This means that the conserved quantity $\gamma$ we computed above can be interpreted as the usual Lorentz boost factor in the dilute limit.
Furthermore, this tells us how a particle's speed depends on its radial position at a given time: $v(r) = r/\tau_0$.

Since we know what the energy-momentum tensor looks like in FRW coordinates, a straightfoward coordinate transformation gives us the Minkowski-space energy-momentum tensor:
\begin{equation}
T_{ab} = \rho dt^2 = \frac{3 t_{\rm c}}{8\pi}\frac{dt^2}{t^3} = \frac{3 t_{\rm c}}{8\pi} \frac{(\tau d\tau -rdr)^2}{(\tau^2-r^2)^{5/2}}~.
\end{equation}
Here we have assumed that we are in the dilute limit, so $a(t) = t$.
We can read off the energy density as the coefficient of $d\tau^2$.
Incidentally, this lets us check one of our earlier calculations by computing the total energy at time $\tau_0$:
\begin{equation}
E = 4\pi \int_0^{r_{\rm max}} \left(\frac{3 t_{\rm c}}{8\pi} \frac{\tau_0^2}{(\tau_0^2-r^2)^{5/2}}\right) r^2dr = \frac{t_{\rm c}}{2}\sinh^3 \xi_{\rm max}~,
\label{eq-totalen}
\end{equation}
where $\tanh \xi_{\rm max} = r_{\rm max}/\tau_0$.
Of course, this matches our result for the Schwarzschild mass $M$.

Now we would like to know the number density of particles.
We know from relativistic mechanics that a particle of mass $m$ and boost $\gamma$ has energy $\gamma m$.
So if we divide the energy density by this factor we will find the number density:
\begin{equation}
n(r) = \frac{1}{m\gamma}\frac{3 t_{\rm c}}{8\pi} \frac{\tau_0^2}{(\tau_0^2-r^2)^{5/2}} = \frac{3 t_{\rm c}}{8\pi m} \frac{\tau_0}{(\tau_0^2-r^2)^{2}}~.
\label{eq-numdensity}
\end{equation}
The total number of particles is
\begin{equation}
N = 4\pi \int_0^{r_{\rm max}} \left(\frac{3 t_{\rm c}}{8\pi m} \frac{\tau_0}{(\tau_0^2-r^2)^{2}}\right) r^2dr = \frac{3t_{\rm c}}{2 m}\int_0^{\xi_{\rm max}}\sinh^2 \xi \,d\xi = \frac{3t_{\rm c}}{8\pi m}V_{\rm c}(\xi_{\rm max})~,
\label{eq-totalnum}
\end{equation}
which is a result we could have anticipated by examining the cosmological solution.

\subsection{Maximizing the entropy}

If one were to attempt to create this collapsing of matter artificially by assembling a ball of dust in Minkowski space, how should the initial positions and velocities of the matter be arranged?  In this section we answer this question by describing the collapsing ball in terms of Minkowski space variables, demonstrating explicitly how a light-sheet with entropy $S \gg A^{3/4}$ can be created.
\begin{figure}[tbp]
\centering
   \includegraphics[width=3 in]{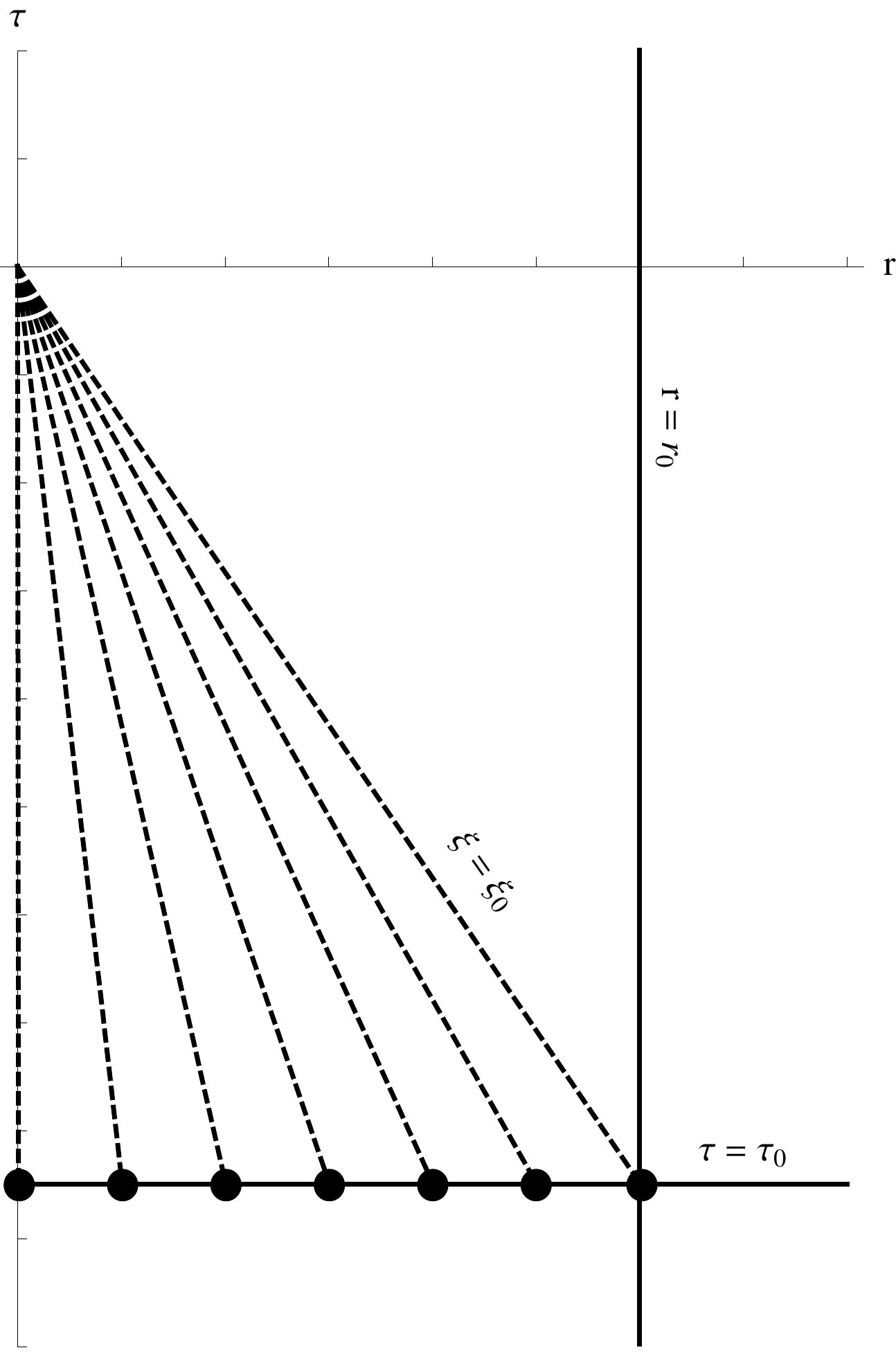}
   \caption{From the Minkowski space point of view, the ball of matter starts at $\tau=\tau_0$ as a dilute gas with a specific radial number density discussed in the text. The velocities of the particles in this must be arranged so that the particles converge to a single point, illustrated in this cartoon by the dashed trajectories all reaching the origin.}
   \label{fig-mink}
\end{figure}

To keep track of our efficiency in this task, we introduce a parameter $\alpha$ encoding the amount by which we beat the naive bound:
\begin{equation}
S = \alpha A^{3/4}~.
\label{eq-alphadef}
\end{equation}
Optimally we would find $\alpha \sim A^{1/4}$, but we will consider ourselves successful if $\alpha>O(1)$.

 To accomplish this feat we will gather a large number $N$ of particles, each of mass $m$, and arrange them in a ball of radius $r_{\rm max}$.
The initial velocities are such that all of the particles converge to a single point in the future, $v(r)\propto r$.
Since we are interested in speeds near the speed of light, it will be more convenient to parametrize the speeds by their Lorentz boost factors $\gamma(r) = (1-v(r)^2)^{-1/2}$.
An additional parameter is the boost factor of the most energetic particles, $\gamma_{\rm max} \equiv \gamma(r_{\rm max})$, which determines how much time passes before the particles collide.
Finally, to ensure that the light-sheet is properly formed according to the description in the previous sections, the number density $n(r)$ must follow Eq.~(\ref{eq-numdensity}), $n(r) \propto \gamma(r)^4$~.

We know from the previous section that the ball of dust will evolve like a portion of an open FRW universe.
To maximize the entropy in our light-sheets, we need to make sure the open nature of the geometry is visible.
This means we must have $\gamma_{\rm max} \gg 1$.
In the language of the previous section, $\gamma_{\rm max} = \cosh \xi_{\rm max}$, so we will freely approximate $\sinh \xi_{\rm max} \approx \cosh \xi_{\rm max} \approx \exp (\xi_{\rm max}) /2$ throughout the following discussion.
To make contact with the formulas of the previous section, we notice that the derived quantity $t_{\rm c}$ is defined by Eq.~(\ref{eq-totalnum}) to be
\begin{equation}
t_{\rm c} = \frac{8\pi Nm}{3 V_{\rm c}(\xi_{\rm max})}\sim~ \frac{Nm}{\gamma_{\rm max}^2}~.
\end{equation}
For the light-sheets which extend to FRW time $t= t_{\rm c}$ inside the ball of matter, the entropy they see is related to their area according to Eq.~(\ref{eq-samatter}) by
\begin{equation}
S \sim (m t_{\rm c})^{-1} A \sim \frac{\gamma_{\rm max}^2}{Nm^2} A~.
\end{equation}
Combining this with Eq.~(\ref{eq-alphadef}) we learn that
\begin{equation}
\alpha \sim \frac{\gamma_{\rm max}^{3/2}}{N^{3/4}m^{3/2}} S^{1/4}~.
\end{equation}
With $\gamma_{\rm max}$, $N$, and $m$ fixed, the entropy $S$ on the light-sheet depends on which light-sheet we choose to look at.
Here we learn that the light-sheet with the best chance of beating $A^{3/4}$ is the one that sees the most entropy.
The maximum entropy any light-sheet can see is $S\sim N$, and there is a light-sheet starting on the apparent horizon for which this is true:
Simply take the light-sheet which reaches $\xi=\xi_{\rm max}$ at the time $t=t_{\rm c}$.
With this optimal choice, we have
\begin{equation}
\alpha \sim \left(\frac{\gamma_{\rm max}^3}{Nm^3}\right)^{1/2}~,
\end{equation}
which can be made large, as we will now show.
Recall from Sec.~(\ref{sec-matter}) that $m>t_{\rm c}^{-1/2}$ is required for our entropy counting to be accurate, and saturating this inequality gives the largest possible $S/A$.
Setting $m\sim t_{\rm c}^{-1/2}$ implies
\begin{equation}
\gamma_{\rm max}^2 \sim Nm^3~,
\end{equation}
and hence
\begin{equation}
\alpha \sim (Nm^3)^{1/4} \sim \gamma_{\rm max}^{1/2}~.
\end{equation}
Since $\gamma_{\rm max}>O(1)$ throughout this discussion, there is no obstruction to getting $\alpha> O(1)$.

\subsection{Example}

It is interesting to consider the possibility of actually creating such a ball of matter out of the materials found in our own universe.
One constraint on our ability to perform this experiment is the energy cost.
The total energy required is, from Eq.~(\ref{eq-totalen}),
\begin{equation}
E \sim N m \gamma_{\rm max} \sim  m^{-2}\alpha^6~.
\end{equation}
The most plentiful element in our universe is hydrogen, so we will attempt to build our dust ball from hydrogen molecules.
The relevant mass, then, is the proton mass, $m\sim 10^{-19}$ in planck units.
The energy we have available to us (from processing all of the stars in our horizon, for instance) is $E\sim 10^{58}$.
This implies the largest $\alpha$ we can manage is
\begin{equation}
\alpha \sim 10^{3.3}~.
\end{equation}
(Not all of the stars can be processed for energy; some hydrogen must remain to form the dust ball itself.
The rest mass of the dust ball is only a small fraction $\alpha^{-2}$ of the total energy, though, so the number of stars we need to set aside for this purpose is negligible.)

This represents a significant improvement over the naive $\alpha \sim 1$, and it is somewhat remarkable that we can accomplish the task even in principle given that our starting material is so far from the optimal $m\sim 1$.  (However, we make no attempt to analyze the stability of this solution or to quantify the extent of fine-tuning in the initial conditions.)

\section{Saturating the entropy bound in a causal diamond}
\label{sec-cd}

We have found many examples of light-sheets with $S\gg A^{3/4}$.  This rules out the naive bound $S\lesssim A^{3/4}$ as a general bound on the entropy of matter.  However, perhaps the bound may still hold under certain restrictions.  It is interesting, from this point of view, that we have found no light-sheet with  $S\gg A^{3/4}$ which {\em also\/} lies within a causal diamond.

The causal diamond for a worldline~\cite{Bou00a} is the region which can both send and receive signals from that worldline, so it is the largest region an observer following that worldline can probe. We showed in sections Sections~\ref{sec-patch} and \ref{sec-ball} that the entropy bound can be saturated on light-sheets that lie within a past {\em or} future light-cone.  But none of them lies within {\em both\/} the past and the future of a single worldline.  That is, they do not fit within the causal diamond.  This implies that a single observer cannot both set up the system and then observe the entropy.  In this section, however, we will show that this, too, can be arranged.

We will give three examples. First, we modify our cosmological examples by replacing the singular Big Bang with a nonsingular bubble nucleation event. This allows us to extend worldlines farther back into the past, enlarging the causal diamond so that the light-sheets of Sec.~\ref{sec-patch} now fit within the causal diamond. Then we give two noncosmological examples: the slow feeding of a black hole, and a shell of dust in Anti-de Sitter space.

\subsection{Coleman-DeLuccia decay}
\label{sec-cdl}

There is a simple, well-motivated modification of our cosmological examples which allows the light-sheets of interest to fit inside one causal diamond. Suppose that instead of beginning with a singular Big Bang, as we have been assuming, an open FRW universe begins by the formation of a Coleman-DeLuccia (CDL) bubble within a false vacuum of larger, positive cosmologica constant. This is a natural assumption in the context of an eternally inflating multiverse. In particular, our own Big Bang may well have been a bubble nucleation.

\begin{figure}[tbp]
\centering
   \includegraphics[width=2 in]{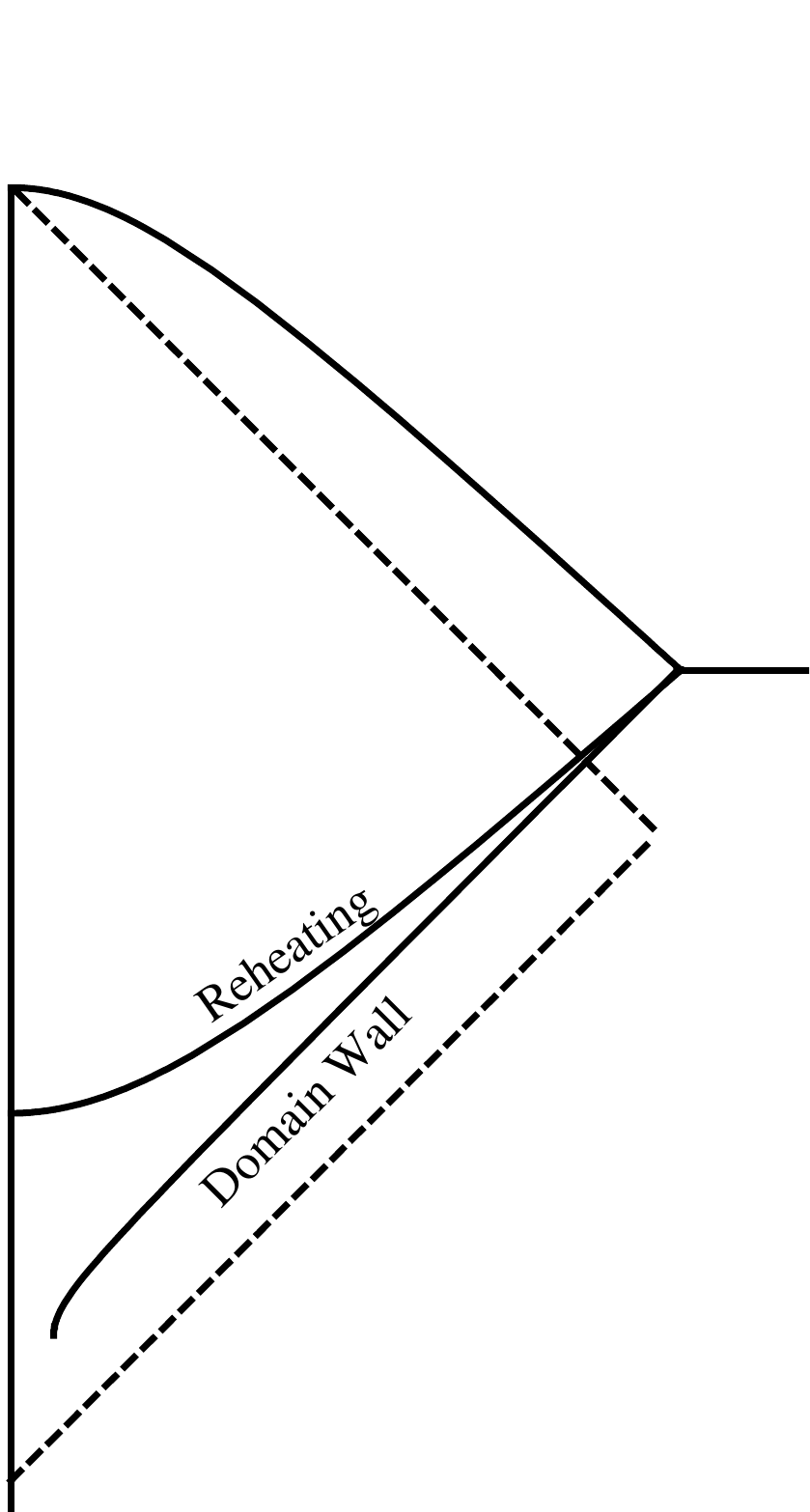}
   \caption{By replacing the Big Bang singularity of the radiation dominated open FRW universe with a bubble nucleation event in a false parent vacuum, the light-sheets shown in Fig.~2 can be made to fit into the causal diamond of a single observer (dashed line).}
   \label{fig-nucleation}
\end{figure}

The spacetime resulting from a CDL bubble is nonsingular, but it is not quite classical because the nucleation is a quantum process. To make it classical, we can imagine classically forming a critical bubble of true vacuum by locally exciting the tunneling field. Once the critical bubble of true vacuum has formed, the expansion of the bubble into the false vacuum and the open FRW universe within the bubble evolve classically (see Fig.~\ref{fig-nucleation}).

The causal diamond of a worldline which begins in the false vacuum is bounded, within the open FRW universe, by the backward light-cone of the observer. In other words, the causal diamond is equivalent to the causal patch within the open FRW universe, and we showed in Sec.~\ref{sec-patch} that the entropy bound is saturated on light-sheets within one causal patch.  Of course, an actual observer is unlikely to survive the transition between vacua.  But this example demonstrates that there is no obstruction to fitting the light-sheets we found earlier into a single causal diamond.

\subsection{Shell of dust in AdS}
\label{sec-adsshell}

We will now show that a shell of matter in Anti-de~Sitter space gives rise to light-sheets with entropy $S\gg A^{3/4}$.  This example is closely related to the ``normal'' light-sheets discussed in Sec.~\ref{sec-nl}: by keeping a portion of an open $\Lambda<0$ FRW universe, one would obtain a ball of dust collapsing to form a Schwarzschild-AdS black hole.  After removing part of the interior of the ball, one obtains a shell of dust instead, which makes the relevant light-sheets fit within a causal diamond.  Here we will simply describe the resulting shell.

Empty AdS space can be written in the ``cosmological'' coordinates
\begin{equation}
ds^2 = - dt^2 + a^2(\tau) (d \chi^2 + \sinh^2 \chi d \Omega_2^2)
\label{eq-adsmetric}
\end{equation}
where $a(t) = t_\Lambda \sin (t/ t_\Lambda)$. We now add
some matter, but for convenience we add a low density of matter so
that we can continue to use the vacuum metric. Worldlines with
constant values of the spatial coordinates in this metric are
geodesics, so if we add dust at rest it will remain at rest in the
probe approximation. 

Suppose we add a shell of matter which is spherically symmetric and
extends from some comoving radius $\chi_{\rm min}$ out to $\chi_{\rm
  max}$. We will find that the most interesting shells for our
purposes are at large radius $\chi \gg 1$ and extend over one
curvature radius, $\Delta \chi \sim 1$. Making a thicker shell does
not add much entropy, because in an open geometry an order one
fraction of the entropy is in the outermost curvature radius.

We add the matter so that it is a small perturbation at the
``turnaround'' time $t = t_\Lambda \pi/2$ when the scale factor is
maximal. For the matter to be a small perturbation, the density must be small
compared to the vacuum energy,
\begin{equation}
\rho_m \ll {1 \over t_\Lambda^2 }
\label{eq-densitybound}
\end{equation}
Because the matter is following geodesics that converge, as the
system evolves the probe approximation will no longer be good. 

As another check of the probe approximation, one could check that the
matter is well outside its Schwarzschild radius. For this, it is more
convenient to use the global coordinates for AdS where the timelike
Killing vector is manifest. For the large radius shells of interest,
this criterion leads to the same condition \eqref{eq-densitybound}.

The spacetime in the vacuum region inside the shell is unperturbed
AdS, while the metric within the shell is approximated by
\eqref{eq-adsmetric} only as long as the matter density is small
compared to the vacuum energy. The spacetime outside the shell is a
piece of an AdS-Schwarzschild black hole. An approximate conformal
diagram is shown in the figure.

\begin{figure}[ht]
\centering
\subfigure[]{
\includegraphics[scale=.7]{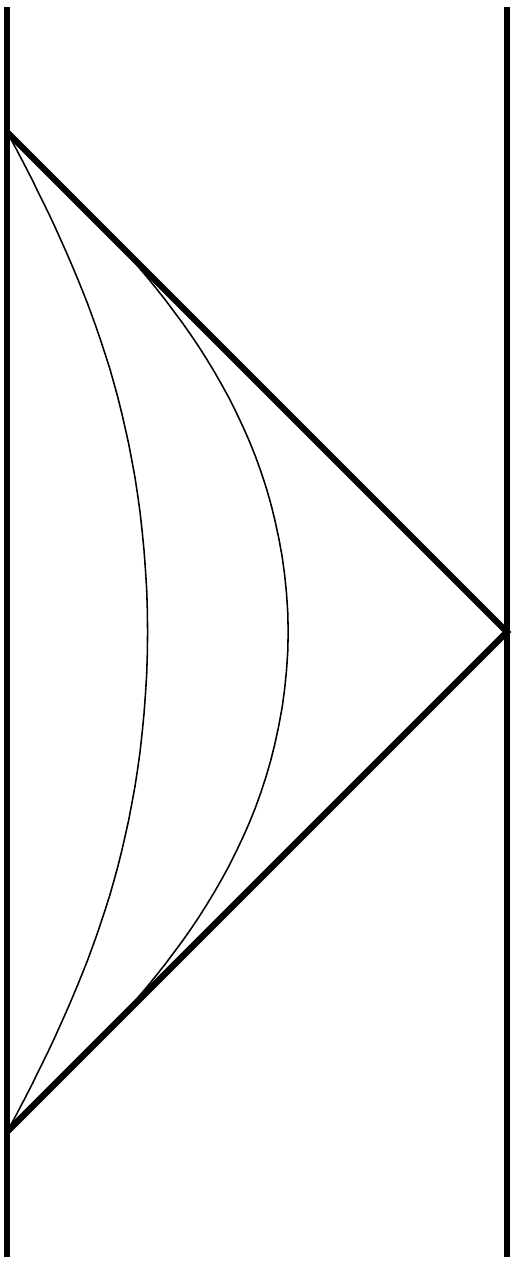}
}
\hspace{1in}
\subfigure[]{
\includegraphics[scale=.7]{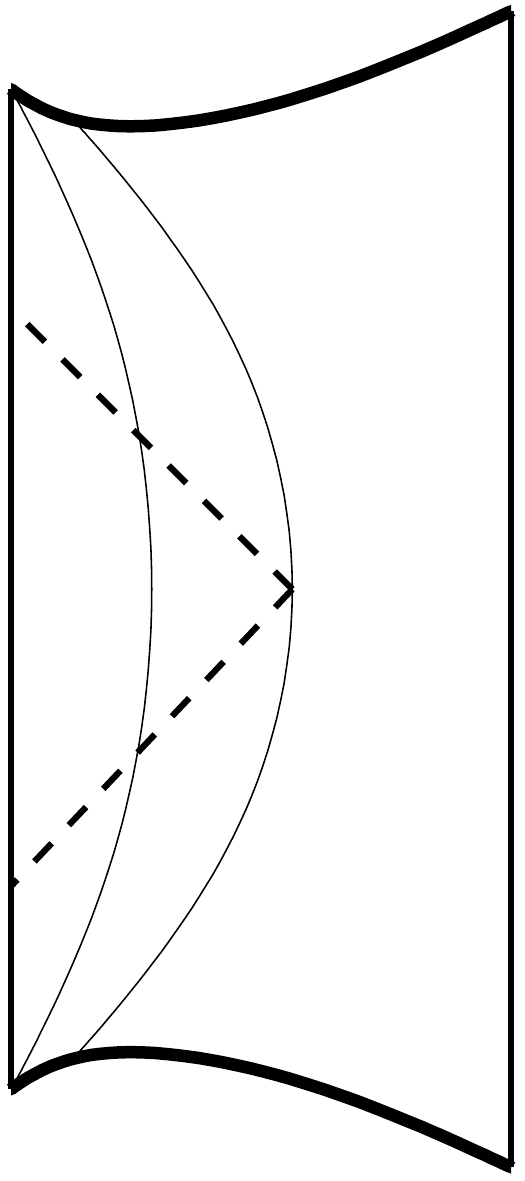}
}
\label{fig-shell}
\caption{A collapsing shell of dust in the probe approximation (left)
  and including backreaction (right). The dashed lines are two light-sheets that contain the entire shell.  They fit inside a single causal diamond.}
\end{figure}

Consider now the ingoing light-sheets- one past directed and one future
directed- that begin at the outer edge of
the shell at the turnaround time. The system is time symmetric about
the turnaround time, so these light-sheets are identical and we discuss
the future-directed one. We are interested in whether such a
light-sheet gets to the origin so that all of the matter is contained
in a single observer's causal diamond, and also the ratio $S/A$.

The radial null rays satisfy
\begin{equation}
d \chi = {dt \over a(t)}
\end{equation}
It is more convenient to write $\Delta \chi$ in terms of the scale
factor. Using the FRW equation
\begin{equation}
\dot a^2 = 1- {a^2 \over t_\Lambda^2}
\end{equation}
gives
\begin{equation}
d \chi = {da \over a \sqrt{1 - a^2/t_\Lambda^2}}
\end{equation}
which integrates to
\begin{equation}
\Delta \chi = \cosh^{-1} \left( t_\Lambda \over a \right)
\label{eq-dchi}
\end{equation}
We want the light-sheet to escape from the shell into the interior
vacuum region before we start to mistrust the probe
approximation. \eqref{eq-dchi} shows that the light ray travels one curvature radius, $\Delta
\chi = 1$, in the time it takes the scale factor to decrease to a few
times smaller than $t_\Lambda$. Because dust evolves as $\rho_m \sim
a^{-3}$, the density increases by a factor of $({\rm a\ few})^3$. Assuming we
started with a matter density comfortably less than the vacuum energy,
the probe
approximation does not break down before the light-sheet of interest
exits the matter shell.

Having escaped into the interior, the light-sheet is guaranteed to reach
the origin before a singularity forms. The metric inside the shell is
vacuum AdS by construction, and the innermost matter particles are following 
timelike geodesics, so the light rays get to the origin before the
metric is affected by the incoming shell of matter. (The proper time
between the arrival of the light-sheet and the arrival of the matter
will be quite small for the shells of interest at large $\chi$.)

Now it remains to compute $S/A$ for this light-sheet. If the matter
is nearly relativistic, then the entropy density  is related to the
energy density,
\begin{equation}
s \sim \rho_m^{3/4}
\end{equation}
The entropy passing through the light-sheet is bounded below by the
entropy on the spatial slice at the turnaround time, which is
\begin{equation}
S = s V = \rho_m^{3/4} A t_\Lambda
\end{equation}
since the shell is one curvature radius thick, and at the turnaround
$a = t_\Lambda$.

To be in the probe approximation we need $\rho_m < \rho_\Lambda$. Therefore the ratio is 
\begin{equation}
{S \over A} \sim \rho_m^{3/4} t_\Lambda \lesssim t_\Lambda^{-1/2}
\end{equation}
Thus we have constructed light-sheets causally accessible to a single
observer on which the entropy is proportional to the area for
arbitrarily large $S$.  By choosing $A\gg \Lambda^{-1}$, we obtain light-sheets with $S\gg A^{3/4}$ which are contained within a causal diamond.

\subsection{Slow feeding of a black hole}
\label{sec-feed}

So far we have analyzed cosmologies and portions thereof.  We will now show the horizon of a black hole can be a light-sheet with $S\gg A^{3/4}$.  

\begin{figure}[tbp]
\centering
   \includegraphics[width=3 in]{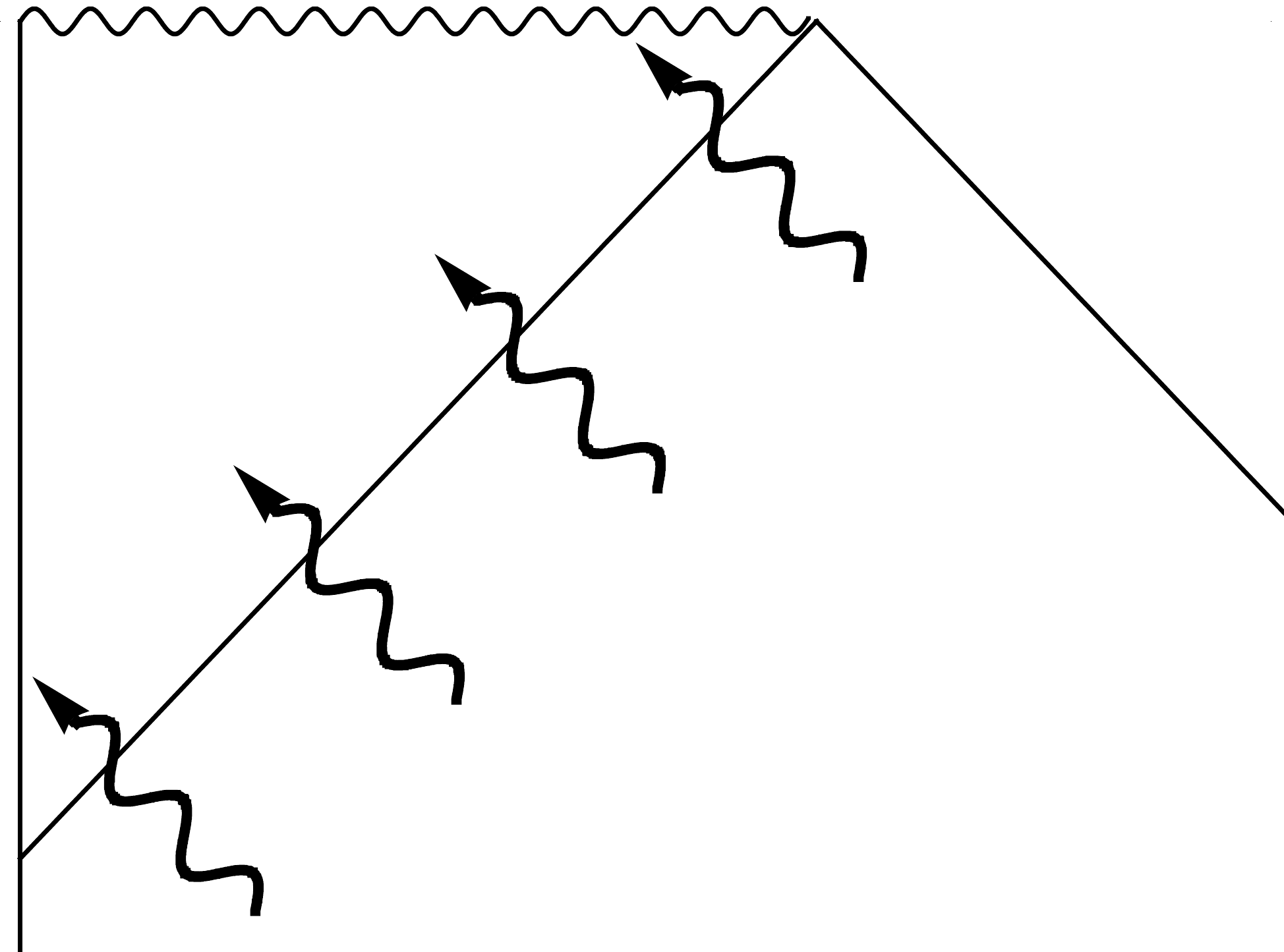}
   \caption{Doubling the radius $R$ of a Schwarzschild black hole by adding massless quanta of wavelength $\lambda\leq R$, one-by-one.  The horizon (diagonal line) can be viewed as a past-directed ingoing light-sheet starting at its largest area (the final area of the black hole, near the top right corner).  The entropy passing through this light-sheet satisfies $S/A\sim \lambda/R$.  This ratio can approach unity, saturating the covariant bound and exceeding the naive bound $S\lesssim A^{3/4}$.}
   \label{fig-slowfeed}
\end{figure}

Consider a black hole of radius $R$; it will not matter how this black hole was created.  We will add individual massless quanta of wavelength $\lambda$ to the black hole, one by one. Each quantum carries energy $\lambda^{-1}$, so the energy and radius of the black hole will have doubled after about $R\lambda$ quanta have been sent across the horizon.    The horizon will now have cross-sectional surfaces of area $4\pi (2R)^2$.  Any such surface admits a past-directed ingoing light-sheet, which contains the portion of the black hole event horizon across which the $R\lambda$ quanta have travelled (see Fig.~\ref{fig-slowfeed}).  Thus, the entropy on the light-sheet is of order $R\lambda$.  The area is of order $R^2$, so 
\begin{equation}
\frac{S}{A}\sim \frac{\lambda}{R}
\end{equation}
Naively, the covariant bound can be violated in this process by choosing $\lambda\gg R$.  However, such large quanta will not be absorbed by the black hole.  The process we have described can be carried out only if $\lambda\leq R$.  Therefore, the covariant bound is satisfied.  (This had better be the case, since the covariant bound reduces to the generalized second law of thermodynamics in the special case of light-sheets that coincide with event horizons~\cite{FMW}.)

Nothing prevents us from making the wavelength nearly as large as the black hole.  In this limit, $\lambda\to R$, the covariant bound is approximately saturated.   More generally, as long as $\lambda\gg R^{1/2}$, the stronger bound $S\lesssim A^{3/4}$ is violated.  (An analogous process can be described for the de~Sitter event horizon.)

Note that the black hole event horizon is a portion of the boundary of the causal diamond of an observer at fixed radius outside the black hole.  Thus, our example shows that light-sheets with $S\gg A^{3/4}$ can exist within a causal diamond.  Indeed, unlike the previous example, this situation can be set up in a laboratory independently of initial conditions, i.e., without relying on the existence of positive vacuum energy and a lower-energy true vacuum.

As originally formulated~\cite{CEB1}, the covariant bound applies to situations in which corrections due to quantum gravitational effects are negligible.  It does not apply in regimes with Planck-scale curvatures, nor does it apply in regimes where the backreaction from Hawking radiation becomes a large correction to the geometry.  This occurs on the black hole event horizon if we follow a light-sheet for a time of order $R^3$, the time it takes the black hole to emit a significant fraction of its mass~\cite{Low99,Bou99d,RMP}.  (Extensions of the covariant bound to this regime have been considered~\cite{StrTho03}.)  In our example, the time it takes to throw in $R\lambda$ quanta of wavelength $\lambda$, one by one, is of order $R\lambda^2$.  Thus, in the regime $\lambda\ll R$, we are operating well within the regime of validity of the covariant bound.  In the saturation limit, $\lambda\to R$, we approach the edge of the regime of validity.  However, corrections will be no larger than factors of order unity, which are neglected here in any case.

It is interesting that in all of our examples, the edge of the regime of validity of the covariant bound is approached in the saturation limit, though in two very different ways. In all previous sections, it arises because most of the entropy on light-sheets with $S\sim A$ lies in a near-Planckian regime, even though $A$ may be large.   In the present section, the breakdown of the bound comes from the fact that the backreaction of Hawking radiation is becoming important.

\section{A conjecture concerning the bound $S\lesssim A^{3/4}$}
\label{sec-conjecture}

We have presented a number of counterexamples to the bound
$S \lesssim A^{3/4}$.  Therefore, this bound cannot be universally valid.  It remains possible that this bound is valid in a more restrictive setting.  This would be of interest as long as the regime of validity can be easily and sharply defined.  As an example, let us consider the following\\[1ex]
{\bf Conjecture: Static, weakly gravitating systems satisfy the
  stronger bound $\mathbf{S < A^{3/4}}$.}\\[1ex]
We have no deep reason for believing the above conjecture, but we have not yet been able to find counterexamples.

In Sec.~\ref{sec-adsshell} we studied a large shell of matter that collapses under geodesic motion in Anti-de~Sitter space.  One might think that it would not be difficult to stabilize this shell, i.e., to prevent its collapse. This would render the system static and thus turn it into a counterexample to the above conjecture.

To stabilize the shell,  we will attempt to build a wall to keep the particles out at large radius.  Consider a thin spherical wall of radius $r_0$, located just inside
the shell of matter.  We require that this wall be constructed from
matter that satisfies the dominant energy condition (roughly, the requirement that the speed of sound not exceed the speed of light).  This limits the
amount of external pressure, $p$, that can be resisted by a wall of a given
mass per unit area, $\varsigma$~\cite{Bou03a}:
\begin{equation}
\varsigma\geq |p/K|~,
\label{eq-vpk}
\end{equation}
where $K$ is the trace of the extrinsic curvature of the wall worldvolume.  For $r_0\gg t_\Lambda$, we have $K\approx 1/t_\Lambda$. The wall has to
at least hold up its own mass. But a wall element of mass $\varsigma\, dA$ sitting at constant $r$ has proper acceleration $1/t_\Lambda$; in other words, it feels a
gravitational force $dF=\varsigma \, dA/t_\Lambda$.  The pressure is the force per unit area, so gravity exerts a pressure 
\begin{equation}
p = {\varsigma \over t_\Lambda}
\end{equation}
Using the bound on the energy density \eqref{eq-vpk}, we see that an ideal wall saturating \eqref{eq-vpk} can just barely hold itself up. It cannot support any additional matter.

We conclude that a conventional wall cannot prevent a large shell of matter in AdS space cannot be prevented from collapsing. Stabilization may still be possible using D-branes or fluxes, or by giving the particles angular momentum or charge; we will not investigate these possibilities here.  We are aware of one way to stabilize a gas of particles at large radius in AdS: by building a large AdS black hole. The thermal atmosphere of the black hole, even neglecting the divergent near-horizon contribution, contains enough entropy to saturate the entropy bound. But the above conjecture does not apply to this system because it is not weakly gravitating.  (For the same reason it does not apply to the example in Sec.~\ref{sec-feed}.)

It will be interesting to find counterexamples or further support for the above conjecture, or to establish other simple sufficient conditions for the bound $S\lesssim A^{3/4}$. 

\acknowledgments
We are grateful to L.~Susskind for discussions.  This work was supported by the Berkeley Center for Theoretical Physics, by the National Science Foundation (award number 0855653), by the Institute for the Physics and Mathematics of the Universe, by fqxi grant RFP2-08-06, and by the US Department of Energy under Contract DE-AC02-05CH11231.

\appendix
\section{Saturating the generalized covariant entropy bound}
\label{sec-geb}

In this appendix, we study the generalized covariant entropy bound~\cite{FMW},
\begin{equation}
S\leq \frac{\Delta A}{4}~,
\end{equation}
where $\Delta A = A_0 - A $ is the difference in areas between the initial area of a light-sheet and its final area.  The final area is non-zero if the light-sheet is terminated prematurely.  We will consider past-directed outgoing light-sheets in open, radiation-dominated FRW universes, as we did in Sec.~\ref{sec-main}.  There, we extended the light-sheets all the way to the Big Bang, or to the beginning of curvature domination, $t_{\rm c}$ (which gave the same entropy up to factors of order unity).  Here, we will  prematurely terminate our light-sheets at $\xi(a)$ with $a\gg t_{\rm c}$.  We will show that the entropy on such light-sheets can saturate the generalized covariant entropy bound, up to a tunable factor $t_{\rm c}^{-1/2}$.   

For simplicity, we will focus on the case $\Lambda=0$.  We consider past-directed outgoing light-sheets with $a_0 \gg a \gg t_{\rm c}$.  Recall that the initial surface must be outside the apparent horizon,
\begin{equation}
 t_{\rm c}\sinh \xi_0 > a_0~,
 \label{eq-ahcond}
\end{equation}
which implies $\xi(a) \gg 1$.
To $O(t_{\rm c}^2/a^2)$, the comoving radius is
\begin{equation}
\xi(a) = \xi_0 + \sinh^{-1}(a_0/t_{\rm c})-\sinh^{-1}(a/t_{\rm c}) \approx  \xi_0 + \log(a_0/a) -\frac{t_{\rm c}^2}{4a^2}~.
\end{equation}
We have neglected terms of $O(t_{\rm c}^4/a^4)$ and $O(t_{\rm c}^2/a_0^2)$.
As we will shortly see, the area would not change at all without the $O(t_{\rm c}^2/a^2)$ term we have included.
One might worry that we will lose too much accuracy in approximating $\sinh \xi \approx \exp(\xi)/2$ when computing the area since we needed to keep a higher order term in the $\xi(a)$ expansion.
To be safe, we will keep the next order term beyond the exponential in the expansion of $\sinh\xi$ until we can verify that it is unnecessary.
For the area, we have
\begin{align}
A(a) &= 4\pi a^2 \sinh^2 \xi(a)= \pi a^2 e^{2\xi(a)}\left(1-e^{-2 \xi(a)}\right)^2\\ 
&\approx \pi a^2e^{2\xi(a)}-2\pi a^2\\
&\approx  \pi a_0^2\left(1-\frac{t_{\rm c}^2}{2a^2} \right)e^{2\xi_0}-2\pi a^2 \\
&= \pi a_0^2e^{2\xi_0} - \frac{\pi t_{\rm c}^2a_0^2}{2a^2}e^{2\xi_0}-2\pi a^2~.
\end{align}
Subtracting this from $A_0=A(a_0)$ and dropping a term of $O(a^2)$ we get
\begin{equation}
\Delta A = A_0-A(a) \approx \frac{\pi t_{\rm c}^2a_0^2}{2a^2}e^{2\xi_0}-2\pi a_0^2~.
\end{equation}
Finally, from Eq.~(\ref{eq-ahcond}), we have $t_{\rm c} \exp \xi_0 > a_0 \gg a$, so we can neglect the second term and reach our final conclusion:
\begin{equation}
\Delta A \approx  \frac{\pi t_{\rm c}^2a_0^2}{2a^2}e^{2\xi_0}~.
\end{equation} 
Now we will compute the entropy on the terminated light-sheet.
In the regime $a_0 \gg a \gg t_{\rm c}$, 
\begin{equation}
S = t_{\rm c}^{3/2}\left(V_{\rm c}(\xi(a)) - V_{\rm c}(\xi_0) \right) \approx \frac{\pi t_{\rm c}^{3/2}a_0^2}{2a^2}e^{2\xi_0}~.
\end{equation}
So the entropy on the terminated light-sheets saturates the bound $S\leq \Delta A/4$, up to a factor of order $t_{\rm c}^{-1/2}$, which can be chosen to approach unity.

\bibliographystyle{utcaps}
\bibliography{all}

\providecommand{\href}[2]{#2}\begingroup\raggedright\begin{thebibliography}{10}

\bibitem{CEB1}
R.~Bousso, ``A covariant entropy conjecture,'' {\em JHEP} {\bf 07} (1999)  004,
\href{http://arxiv.org/abs/hep-th/9905177}{{\tt hep-th/9905177}}.

\bibitem{RMP}
R.~Bousso, ``The holographic principle,'' {\em Rev. Mod. Phys.} {\bf 74} (2002)
   825,
\href{http://arXiv.org/abs/hep-th/0203101}{{\tt hep-th/0203101}}.

\bibitem{FisSus98}
W.~Fischler and L.~Susskind, ``Holography and cosmology,''
  \href{http://arxiv.org/abs/hep-th/9806039}{{\tt hep-th/9806039}}.

\bibitem{KalLin99}
N.~Kaloper and A.~Linde, ``Cosmology vs. holography,''
\href{http://arxiv.org/abs/hep-th/9904120}{{\tt hep-th/9904120}}.

\bibitem{FMW}
E.~E. Flanagan, D.~Marolf, and R.~M. Wald, ``Proof of Classical Versions of the
  {B}ousso Entropy Bound and of the {G}eneralized {S}econd {L}aw,'' {\em Phys.
  Rev. D} {\bf 62} (2000)  084035,
\href{http://arxiv.org/abs/hep-th/9908070}{{\tt hep-th/9908070}}.

\bibitem{BouFla03}
R.~Bousso, E.~E. Flanagan, and D.~Marolf, ``Simple sufficient conditions for
  the generalized covariant entropy bound,'' {\em Phys. Rev. D} {\bf 68} (2003)
   064001,
\href{http://arxiv.org/abs/hep-th/0305149}{{\tt hep-th/0305149}}.

\bibitem{CEB2}
R.~Bousso, ``Holography in general space-times,'' {\em JHEP} {\bf 06} (1999)
  028,
\href{http://arxiv.org/abs/hep-th/9906022}{{\tt hep-th/9906022}}.

\bibitem{Mal97}
J.~Maldacena, ``The Large {$N$} limit of superconformal field theories and
  supergravity,'' {\em Adv. Theor. Math. Phys.} {\bf 2} (1998)  231,
  \href{http://arxiv.org/abs/hep-th/9711200}{{\tt hep-th/9711200}}.

\bibitem{SusWit98}
L.~Susskind and E.~Witten, ``The holographic bound in {A}nti-de~{S}itter
  space,'' \href{http://arxiv.org/abs/{h}ep-th/9805114}{{\tt
  {h}ep-th/9805114}}.

\bibitem{SusTho93}
L.~Susskind, L.~Thorlacius, and J.~Uglum, ``The Stretched horizon and black
  hole complementarity,'' {\em Phys. Rev. D} {\bf 48} (1993)  3743,
\href{http://arxiv.org/abs/hep-th/9306069}{{\tt hep-th/9306069}}.

\bibitem{HsuRee07}
S.~D.~H. Hsu and D.~Reeb, ``{Black hole entropy, curved space and monsters},''
  \href{http://dx.doi.org/10.1016/j.physletb.2007.09.021}{{\em Phys. Lett.}
  {\bf B658} (2008)  244--248},
\href{http://arxiv.org/abs/0706.3239}{{\tt arXiv:0706.3239 [hep-th]}}.

\bibitem{HsuRee09}
S.~D.~H. Hsu and D.~Reeb, ``{Monsters, black holes and the statistical
  mechanics of gravity},''
  \href{http://dx.doi.org/10.1142/S0217732309031624}{{\em Mod. Phys. Lett.}
  {\bf A24} (2009)  1875--1887},
\href{http://arxiv.org/abs/0908.1265}{{\tt arXiv:0908.1265 [gr-qc]}}.

\bibitem{BouLei09}
R.~Bousso and S.~Leichenauer, ``{Predictions from Star Formation in the
  Multiverse},''
\href{http://arxiv.org/abs/0907.4917}{{\tt arXiv:0907.4917 [hep-th]}}.

\bibitem{Bou06}
R.~Bousso, ``Holographic probabilities in eternal inflation,'' {\em Phys. Rev.
  Lett.} {\bf 97} (2006)  191302,
\href{http://arxiv.org/abs/hep-th/0605263}{{\tt hep-th/0605263}}.

\bibitem{BouHar07}
R.~Bousso, R.~Harnik, G.~D. Kribs, and G.~Perez, ``Predicting the cosmological
  constant from the causal entropic principle,'' {\em Phys. Rev. D} {\bf 76}
  (2007)  043513,
\href{http://arxiv.org/abs/hep-th/0702115}{{\tt hep-th/0702115}}.

\bibitem{LinVan09}
A.~Linde and V.~Vanchurin, ``{How many universes are in the multiverse?},''
\href{http://arxiv.org/abs/0910.1589}{{\tt arXiv:0910.1589 [hep-th]}}.

\bibitem{Bou00a}
R.~Bousso, ``Positive vacuum energy and the {N}-bound,'' {\em JHEP} {\bf 11}
  (2000)  038,
\href{http://arxiv.org/abs/hep-th/0010252}{{\tt hep-th/0010252}}.

\bibitem{Low99}
D.~A. Lowe, ``Comments on a covariant entropy conjecture,'' {\em JHEP} {\bf 10}
  (1999)  026,
\href{http://arXiv.org/abs/hep-th/9907062}{{\tt hep-th/9907062}}.

\bibitem{Bou99d}
R.~Bousso, ``The holographic principle for general backgrounds,'' {\em Class.
  Quant. Grav.} {\bf 17} (2000)  997,
\href{http://arxiv.org/abs/hep-th/9911002}{{\tt hep-th/9911002}}.

\bibitem{StrTho03}
A.~Strominger and D.~M. Thompson, ``A quantum {B}ousso bound,''
\href{http://arxiv.org/abs/hep-th/0303067}{{\tt hep-th/0303067}}.

\bibitem{Bou03a}
R.~Bousso, ``Bound states and the {B}ekenstein bound,'' {\em JHEP} {\bf 02}
  (2004)  025,
\href{http://arxiv.org/abs/hep-th/0310148}{{\tt hep-th/0310148}}.

\end{thebibliography}\endgroup

\end{document}